\documentclass[]{article}

\usepackage{amssymb}
\usepackage{amsmath}
\usepackage{amsthm}
\usepackage{amsfonts}
\usepackage{hyperref}
\usepackage{bbm}
\usepackage{xcolor}
\usepackage{tikz}
\usetikzlibrary{positioning}
\usepackage{graphicx}
\usepackage{caption}
\usepackage{subcaption}
\usepackage{booktabs}
\usepackage{array}
\usepackage{paralist}
\usepackage{verbatim}
\usepackage{enumerate}
\usepackage{tcolorbox}
 \usepackage[margin=1in]{geometry}
\usepackage[nolabel, final]{showlabels}
\usepackage{authblk}
\usetikzlibrary{calc}
\newtheorem{heuristic}{Heuristic}
\newtheorem{paradox}{Paradox}

\newtheorem*{mainresult*}{Main result}

\title{The SIS process on Erd\H{o}s-R\'enyi graphs: determining the infected fraction}
\author[1]{O.S. Awolude}
\author[1]{E. Cator}
\author[1]{H. Don}
\affil[1]{\textit{Department of Mathematics, Radboud University, Heyendaalseweg 135, Nijmegen, 6525 AJ, Gelderland, Netherlands}}

\begin{document}
	
	\maketitle
	
	\begin{abstract}
	There are many methods to estimate the quasi-stationary infected fraction of the SIS process on (random) graphs. A challenge is to adequately incorporate correlations, which is especially important in sparse graphs. Methods typically are either significantly biased in sparse graphs, or computationally very demanding already for small network sizes. The former applies to Heterogeneous Mean Field and to the $N$-intertwined Mean Field Approximation, the latter to most higher order approximations.
	
	In this paper we present a new method to determine the infected fraction in sparse graphs, which we test on Erd\H{o}s-R\'enyi graphs. Our method is based on degree-pairs, does take into account correlations and gives accurate estimates. At the same time, computations are very feasible and can easily be done even for large networks.
\end{abstract}
	
	
\section{Introduction}

The Covid-19 virus was first identified in Wuhan (China) in December 2019, when it started to spread very quickly worldwide. In this initial stage, the number of cases was growing exponentially fast and the disease was disrupting society. In the next few years, many people got infected and there were several peaks of infections of the virus and its variants. Right now, it seems the situation is stabilizing. A large part of the world population has been infected at least once and many people have built up immunity. The number of new cases has dropped, but the disease still goes around and is expected to stay forever. The World Health Organization (WHO) shifts its focus from emergency response to long term Covid-19 disease management \cite{WHO}. In this paper, we show how to estimate and explain the behaviour of infectious diseases after stabilization.  

Mathematical models for infectious diseases have been studied over the last decades, with a recently boosted interest due to the Covid-19 outbreak. In this work we focus on the Markovian SIS process (contact process in the mathematical literature) on finite graphs. In this model, individuals are represented by nodes in a graph, which are either healthy or infected. Each infected node heals at rate 1, and it infects each of its healthy neighbors at rate $\tau$. This means that healthy individuals are always susceptible to the disease, like in the endemic Covid-19 phase. It motivates the terminology susceptible-infected-susceptible (SIS). For more background on the SIS process and related models, we refer to \cite{MN00,lig99,DHB13,KMS}.

The SIS process was first introduced by Harris \cite{H74} in 1974. Harris studied the process on the integer lattice $\mathbb{Z}^d$. It has later been studied on many other graphs \cite{L96, Mou93, CM13,P92,Stacey01}, with recent attention to the process on random graphs \cite{BNNS21,CD21,HD20,MouValYao13,MV16}. Random graphs are designed to model social connections within populations, and as such are suitable to model a population in which an infectious disease is spreading. Our focus in this paper is on the SIS process on Erd\H{o}s-R\'enyi random graphs \cite{ER59,G59}. For background on random graphs in general and the Erd\H{o}s-R\'enyi graph in particular, we refer to \cite{ER60,B85,Hofstad17}.

A key characteristic of the SIS model on the Erd\H{o}s-R\'enyi graph (and many other graphs) is the existence of an epidemic threshold. Below this threshold, the infection dies out exponentially fast. Above the epidemic threshold, extinction only happens at a very large time scale, see \cite{CD21,DurLiu88,GMT05}. On finite graphs, ultimately the process will reach the only stable solution in which all individuals are healthy. However, before extinction, the process will be in a \emph{metastable} or \emph{quasi-stationary} state, which is quickly reached \cite{HM18} and corresponds to the Covid-19 endemic phase. 	Our primary goal in this paper is to get a better understanding of this stabilized behaviour of the SIS-process on sparse graphs, for which the Erd\H{o}s-R\'enyi graph is our main test case. We are particularly aiming at determining the infected fraction of the population in the metastable state based on the parameters of the process and on graph characteristics. 

\subsection{Review of models for the infected fraction}

Exact analysis of the SIS process and its metastable state is hard, and only few rigorous results are available, even for quite simple graphs \cite{CM13}. For a general graph given by its adjacency matrix, the probability distribution on the state space can in principle be computed numerically at any time. However, due to the exponential size of the state space, these matrix computations are limited to very small populations. 

This is the reason that a lot of attention has gone to methods for approximating the behaviour of the process, see \cite{PCMV15,KMS} for a discussion of some different approaches. Kephart and White consider the process on regular graphs, and derive a differential equation for the evolution of the infected fraction \cite{KW91}. Wang et al. \cite{WCWF03} had the interesting idea to generalize this to graphs with arbitrary adjacency matrix, but their results were shown to be inaccurate in the metastable regime \cite{MOK09}. This inspired Van Mieghem et al. \cite{MOK09} to introduce the $N$-intertwined Mean Field Approximation (NIMFA), which requires solving a system with `only' $n$ unknowns to approximate the metastable infected fraction ($n$ being the population size). As the authors note, NIMFA overestimates the metastable infected fraction because it ignores correlations between nodes. This is especially problematic if degrees in the graph are small. 

The Heterogeneous Mean Field approximation (HMF) in \cite{PV01} also ignores correlations. In fact, for graphs with exponentially decaying degree distributions (like the Erd\H{o}s-R\'enyi graph), \cite{PV01} proposes to ignore fluctuations in the degree and just use the average. This is the \emph{homogeneous} mean field approximation. Although this is reasonable if the degrees are not too small, it leads to inaccurate estimates for sparse Erd\H{o}s-R\'enyi graphs. HMF is proposed in \cite{PV01} for scale-free graphs, i.e. graphs in which the degree distribution decays polynomially. HMF is a degree-based mean field approach (DBMF), which can also be used for sparse Erd\H{o}s-R\'enyi graphs. Also HMF suffers from overestimation because correlations are ignored.

NIMFA is also known as the individual-based mean field approach (IBMF). There are several ways to include higher order terms and extend this to a pair-based mean field approach (PBMF), also known as pair-quenched mean field (PQMF), see for instance \cite{CM12,MF13,SFCPC19,SRF20}. These methods have improved accuracy at the cost of more computations. In some cases the solutions might be unstable and physically meaningless \cite{CM12}. Another approach is to use the master equation given in \cite{G11}. This is computationally very demanding, even for small networks. In practice, one has to use approximate forms. The cavity master equation (CME) \cite{ADDM17} gives an alternative system of coupled differential equations, based on conditional probabilities rather than expectations of products. In its full form, it is again challenging to solve. By degree-based averaging, the system can be reduced to a feasible size, coined CME-1 in \cite{OML22}. This greatly simplifies the calculations and is quite promising, although the system is not closed and can not be solved without a suitable Ansatz. Higher order CME approximations take a larger neighborhood of each node into account \cite{MM21}. Also here, assumptions to close the system are needed.  See \cite{OML22} for a more detailed comparison of the different higher order methods. In summary, there is always a trade-off between the level of accuracy and the feasibility of computations.

Our main results are listed in Sec.~\ref{sec:results}, after the definitions in \ref{sec:def}, a comparison with the complete graph in \ref{sec:naive}, and a discussion on annealed and quenched estimation in \ref{sec:ann_que}. When degrees in the graph are small, it is important to take the degree distribution and correlations between nodes into account. Section \ref{sec:improved} builds the framework for all our heuristics. Both NIMFA and HMF fit into this framework. Section \ref{sec:bestheuristic} demonstrates how to include correlations, and discusses asymptotics when the degrees go to infinity. Section \ref{sec:numerics5} presents the numerical results. This includes a discussion on subcritical disconnected Erd\H{o}s-R\'enyi graphs. We also show how our methods can be used to estimate correlations and individual infection probabilities. The methods are suitable to be applied beyond Erd\H{o}s-R\'enyi graphs.

\section{Preliminaries and goals}

\subsection{Model definitions}\label{sec:def}

The goal of this paper is to predict and understand the behaviour of the SIS process on Erd\H{o}s-R\'enyi graphs. In particular we are interested in the quasi-stationary behaviour. Which fraction of the population will be infected on average if the process has reached its metastable equilibrium? 

We denote the set of nodes in our Erd\H{o}s-R\'enyi graph by $V=\{1,2,\ldots,n\}$. Each pair of nodes $\{i,j\}$ is connected (notation: $i\sim j$) with probability $p$, independent of all other pairs. This random graph is denoted $G_{n,p}$. The same notation and terminology will be used for realizations of the random graph as well. We typically take $p$ to be a decreasing function of $n$. For instance, if $p=c/n$ for some constant $n$, the average degree does not depend on the size of the population. In this case $c$ has to be greater than 1, otherwise the graph only has small components which do not interact. If $c>1$, there is a unique \emph{giant} component, containing a positive fraction of the population also for $n\to\infty$. The graph then is called supercritical. All Erd\H{o}s-R\'enyi graphs in this paper are assumed to satisfy this criterion. Another threshold is $p=\log(n)/n$, when (almost) all nodes are in the giant component. 

On such an Erd\H{o}s-R\'enyi graph $G=(V,E)$, we define our Markovian SIS process. This process is a continuous-time process, with state space $2^V$. The state represents the set of infected nodes. Suppose $I\subseteq V$ is the set of infected nodes. The transition rates are then defined by
%\begin{widetext}
\begin{align}
&I\to I\setminus \{i\} \text{\ with\  rate\ $\delta$},&i\in I,&\nonumber\\
&I\to I\cup\{j\}\text{\ with\ rate\ }\#\{i\in I:i\sim j\}\cdot \tau,&j\in V\setminus I.&\nonumber
\end{align} 
%\end{widetext}
The healing rate for each infected node is $\delta$. Throughout this paper, we will stick to the convention that $\delta = 1$, without loss of generality. The infection rate of an infected node to each of its healthy neighbors is $\tau$. Also for $\tau$, we will often take a function of $n$.

\begin{figure}
	\includegraphics[width=\columnwidth]{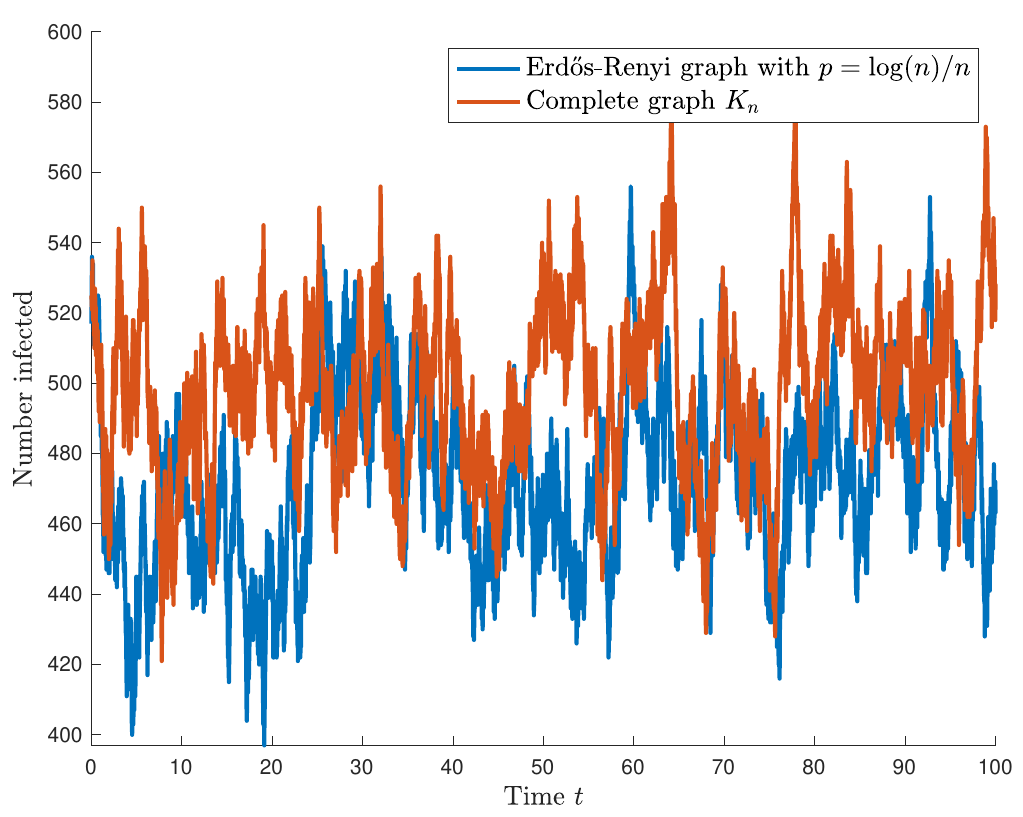}\caption{Number of infected nodes over time for the complete graph and an Erd\H{o}s-R\'enyi graph. In both cases $n=1000$ and the global infection rate $p\tau$ is equal to $2/n$, but the ER graph has a lower equilibrium.}\label{fig:complete_vs_ER}
\end{figure}

For each node $i$, we let $X_i(t)$ denote the status of node $i$ at time $t$, where $X_i(t)$ is either 1 (infected) or 0 (healthy). The total number of infected nodes at time $t$ is denoted by $X(t)$. Furthermore, we let 
\[
\overline X(t) = \frac 1 n\sum_{i=1}^n X_i(t) =\frac{X(t)}{n} 
\]
denote the infected fraction of the population at time $t$. Quasi-stationarity means that the probability distribution of these quantities is (almost) independent of $t$ in a large time window. Such behaviour will happen if the parameters admit a non-trivial state where healings and infections are balanced. For the complete graph, it is known that the quasi-stationary distribution is a well-defined object. This requires taking limits of $n$ and $t$ in the right way, see \cite{ACD23}. Similar behaviour is observed to appear in other graphs, in particular Erd\H{o}s-R\'enyi graphs. Figure \ref{fig:complete_vs_ER} clearly shows that the number of infected individuals fluctuates around some kind of equilibrium. In this paper we aim at predicting and understanding this equilibrium.

\subsection{A naive estimate for the infected fraction}\label{sec:naive}

The SIS process on the complete graph $K_n$ is a birth-death process and is quite well understood. For instance, the time to reach metastability is studied in \cite{HM18}, the average time until extinction in \cite{BM15}. A slight modification of the process by forbidding the transition to the all-healthy state is considered in \cite{CM13}. This gives a very accurate approximation of the metastable distribution of the SIS process, for which exact but somewhat implicit expressions are given. The metastable average infected fraction is $(1-\lambda^{-1})n$ if the infection rate is $\tau = \lambda n^{-1}$ for some constant $\lambda>1$, see \cite{APM22,M11}. Convergence of the metastable infected fraction to a Gaussian with variance $\lambda^{-1}n$ is rigorously proved in \cite{ACD23}. This result can be used for a first naive estimate for the metastable infected fraction in an Erd\H{o}s-R\'enyi graph. 

Consider the SIS process with infection rate $\tau$ on the Erd\H{o}s-R\'enyi graph $G_{n,p}$. An infected node can only infect a healthy node if there is an edge between them. Now assume that the event of two nodes being connected is independent of the event that exactly one of them is infected \footnote{This assumption is reasonable if degrees are not too small. In reality we expect connected nodes to be positively correlated, i.e. if they are connected, it is less likely that exactly one of them is infected. We will come back to this issue later.}. The product $p\tau$ could then be interpreted as the expected infection rate between a random infected and a random healthy node and will be called the \emph{global infection rate} of the process.  Heuristically, the behaviour on an Erd\H{o}s-R\'enyi graph with edge probability $p$ and infection rate $\tau$ should be comparable to the behaviour on the complete graph with the same global infection rate. We therefore take $p\tau$ of order $n^{-1}$ so that we can write $p\tau = \lambda n^{-1}$ for some constant $\lambda>1$. Based on the heuristic that the global infection rate determines the quasi-stationary infected fraction, comparison with the complete graph (which is $G_{n,p}$ with $p=1$) gives the following prediction

\begin{heuristic}\label{heuristic1} Consider the SIS process on an Erd\H{o}s-R\'enyi graph $G_{n,p}$ with global infection rate $p\tau = \lambda n^{-1}$. The quasi-stationary distribution satisfies
	\[
	X\sim N(\mu,\sigma^2),\qquad \text{with}\ \mu = \left(1-\lambda^{-1}\right)n,\ \sigma^2 = n\lambda^{-1}.
	\]
\end{heuristic}
We could make this into a heuristic for the infected fraction, by dividing by $n$. This gives 
\[
\overline X\sim N\left(1-\lambda^{-1},(\lambda n)^{-1}\right),
\] 
so the infected fraction converges to $1-\lambda^{-1}$ if $n\to\infty$. Heuristic \ref{heuristic1} asymptotically gives the correct expectation if the degrees go to infinity ($np\to\infty$). This can be proved using birth-death chains and similar methods as in \cite{CD21}. If $np$ is constant, the expectation in Heuristic \ref{heuristic1} is asymptotically incorrect. The variance is harder to control, we are not aware of any asymptotic results on the variance.

For efficience of simulation, we will mostly restrict to the case $\lambda=2$. We think that other choices give qualitatively similar behaviour. For an illustration of other values, see Figure \ref{fig:Heuristics_Simulation} on page~\pageref{fig:Heuristics_Simulation}.

To assess Heuristic \ref{heuristic1}, Figure \ref{fig:complete_vs_ER} compares a simulation of the process on an Erd\H{o}s-R\'enyi graph with a simulation on $K_n$. At $t=0$, each node is infected with probability $1/2$ independently. Taking the average over time, the process on the complete graph has half of the population infected (this has been rigorously proved in \cite{ACD23}). Heuristic \ref{heuristic1} predicts the same for the Erd\H{o}s-R\'enyi graph. Apparently, the average infected fraction in \emph{this realization} of the Erd\H{o}s-R\'enyi graph is lower, the picture shows an average around 470.

\begin{figure}
	\includegraphics[width=\columnwidth]{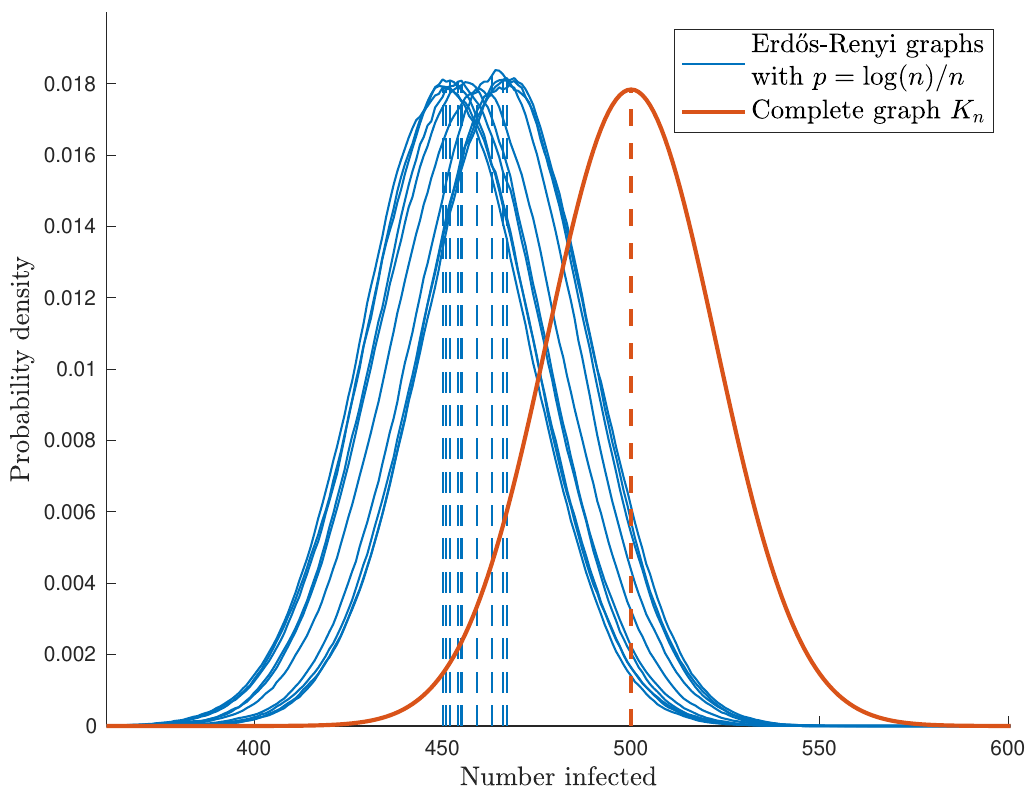}\caption{Probability distribution of the number of infected nodes in the complete graph (red) and in 10 realizations of the Erd\H{o}s-R\'enyi graph (blue), all with $n=1000$ and global infection rate $2/n$. Dashed lines indicate means. }\label{fig:statdistsER}
\end{figure}

It could still be that the heuristic works well if we would average over multiple realizations of the Erd\H{o}s-R\'enyi graph. We therefore repeated the simulation and generated 10 independent realizations of the Erd\H{o}s-R\'enyi graph. On each of these graphs, we simulated the SIS process for $0\leq t\leq 1000$. See Figure \ref{fig:statdistsER} for the (approximated) quasi-stationary distribution for each of the graphs and a comparison with the quasi-stationary distribution of the complete graph.

Based on Figure \ref{fig:statdistsER}, there seems to be a systematic overestimation of the mean in Heuristic \ref{heuristic1}, which does not vanish if we would average over multiple realizations of the Erd\H{o}s-R\'enyi graph. The mean of the quasi-stationary distribution is typically smaller than $n/2$. Apparently Heuristic \ref{heuristic1} is too naive. One reason is that the infected set is not a uniform selection of nodes. This is an essential difference with the complete graph, where transition rates only depend on the size of the infected set, not on the exact selection of infected nodes. In Erd\H{o}s-R\'enyi graphs, nodes with higher degrees are more likely to be in the infected set. Another point is that there are local effects: neighbors tend to be infected or healthy simultaneously. These differences make the model more realistic, but at the same time more complicated. We can not just take averages over all nodes as in the complete graph, but need more subtle estimation methods.

\subsection{Annealed and quenched estimates}\label{sec:ann_que}

When looking at Figure \ref{fig:statdistsER}, we see that the distribution in the Erd\H{o}s-R\'enyi graph is not only clearly different from the complete graph, but also depends on the realization of the graph. For a given graph $G$, we can estimate the metastable distribution by simulating the SIS process on this particular graph. The corresponding random variable depends on $G$ and on the infection rate $\tau$ and is denoted $X(G,\tau)$. Given $G$, the metastable mean infected number is a constant $\mu(G,\tau):=\mathbb{E}[X(G,\tau)]$. The expectation is taken with respect to the randomness of the process, not of the graph. We will call $\mu(G,\tau)$ the \emph{quenched mean} and $\mu(G,\tau)/n$ the \emph{quenched infected fraction}. Each blue curve in Figure \ref{fig:statdistsER} has its own quenched mean. Similarly, we denote the quenched variance by $\sigma^2(G,\tau)$.

If we take the graph to be random, then $\mu(G,\tau)$ becomes a random variable. Now we can take the expectation over the randomness of the graph as well to find the constant $\mathbb{E}[\mu(G,\tau)]$. This is what we will call the \emph{annealed mean}. The annealed mean is independent of the graph realization and can be simulated by generating a bunch of graphs, running the process on each of them and taking the average of their metastable quenched means. For the SIS process on Erd\H{o}s-R\'enyi graphs, the annealed mean is a function of the parameters which we denote by $\mu(n,p,\tau)$.

This gives rise to two different questions concerning estimation of the metastable infected fraction (or other quantities) in $G=G_{n,p}$:
\begin{enumerate}
	\item \textbf{Annealed estimation:} given $n$, $p$ and $\tau$, estimate $\mu(G,\tau)$, without observing the realization of $G$. The estimate will be a function $\hat\mu(n,p,\tau)$.
	\item \textbf{Quenched estimation:} given the realization of $G$, estimate $\mu(G,\tau)$ using the graph information. The estimate will be a function $\hat\mu(G,\tau)$.
\end{enumerate}
When doing quenched estimation, we are estimating the constant $\mu(G,\tau)$ for fixed $G$. The metastable infected fraction is some complicated function of $G$ and $\tau$ which in principle could be determined exactly.  However, when doing an annealed estimation, the realization of $G$ is not known and $\mu(G,\tau)$ is a random variable. This implies that we will always make errors caused by the variation of this random variable. It therefore makes sense to investigate the order of the variance of $\mu(G,\tau)$ when $G$ is random.

\begin{figure}
	\includegraphics[width=\columnwidth]{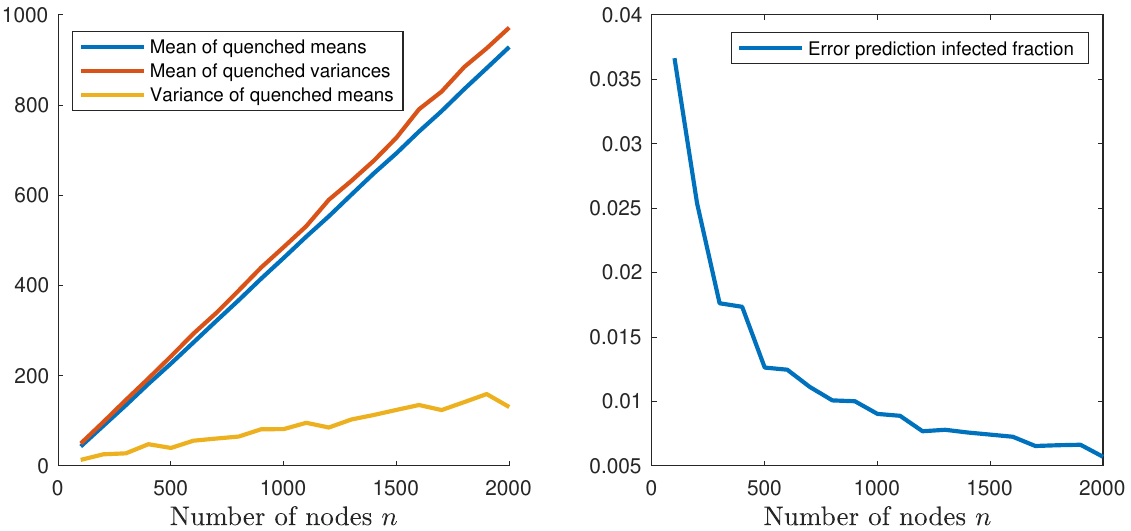}
	\caption{Left: Fixing an Erd\H{o}s-R\'enyi graph ($p=\log(n)/n$, $p\tau = 2/n$), the metastable number of infected has mean (blue) and variance (red) both close to $n/2$. Varying the graph, the means themselves also fluctuate with a variance of order $n$ (yellow). Right: Approximate minimal standard error for any annealed method to estimate a quenched infected fraction.}\label{fig:annealedquenched}
\end{figure}

To do this, we generated for different values of $n$, realizations $G_n^{(1)},\ldots,G_n^{(100)}$ of an Erd\H{o}s-R\'enyi graph $G_{n,p}$. On each of them, we simulated the quenched mean $\mu(G_n^{(k)},\tau)$ and quenched variance $\sigma^2(G_n^{(k)},\tau)$ of the metastable distribution. Then we took averages over $k$  to obtain the mean of quenched means and the mean of quenched variances. Plotting them (Figure \ref{fig:annealedquenched}, left) shows that quenched means and quenched variances both grow linearly in $n$. In fact they are both close to $n/2$, as could be expected by comparing to the complete graph (cf. Heuristic \ref{heuristic1}).

Important is that also the variance of the quenched means appears to grow linearly with $n$. This indicates that annealed estimates for $\mu(G,\tau)$ will always have errors of the same order as fluctuations in the metastable distribution (i.e. the square root of the quenched variance). Therefore, graph information is for all graph sizes of significant importance to make accurate estimates. Taking the square root of the variance of the quenched means and dividing by $n$ gives the standard deviation in annealed estimation of $\mu(G,\tau)$, see right plot in Figure \ref{fig:annealedquenched}. For instance, for $n=1000$, the standard deviation is about 0.01. This implies that any annealed estimation method for the infected fraction will make errors of at least this order.   

\begin{figure}
	\includegraphics[width=\columnwidth]{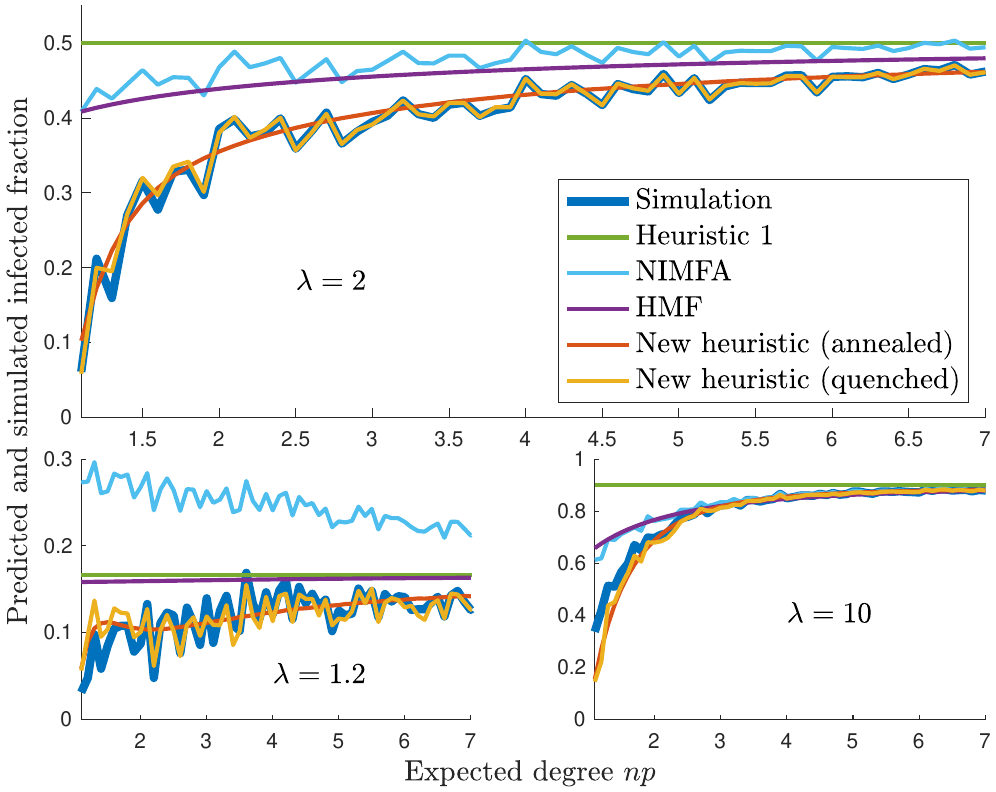}\caption{Simulated infected fraction compared to different estimates. In all cases, $n=1000$. Annealed methods are smooth, quenched methods follow the fluctuations of the simulation.}\label{fig:Heuristics_Simulation}
\end{figure}

This point is once again illustrated in Figure \ref{fig:Heuristics_Simulation}. Here we see simulations of infected fractions in Erd\H{o}s-R\'enyi graphs for different values of $p$ and $\lambda$. Some of these estimates are annealed (corresponding to the smooth curves) and some of them are quenched (the non-smooth curves). The quenched estimates fluctuate much more, but follow the simulation result much better. An annealed estimate can never reproduce these fluctuations, because they are caused by the randomness of the graph. 

Figure \ref{fig:Heuristics_Simulation} also shows that HMF systematically overestimates the infected fraction in these sparse graphs. The same holds for NIMFA, although it clearly follows the random fluctuations of the simulation due to the quenched nature of NIMFA. The new heuristics presented in this paper are much more accurate, both their annealed and quenched versions. 

\subsection{Main results}\label{sec:results}

The main result of this paper is a new method to estimate the infected fraction for connected graphs:

\begin{mainresult*}\label{mainresult} Algorithm (details in Figure \ref{fig:mainresult} on page~\pageref{fig:mainresult}) to estimate the metastable infected fraction $\mu/n$ in a connected graph $G$. 
	\textbf{Input:} Degree sequence of $G$, infection rate. \textbf{Output:} Prediction for $\mu/n$.
\end{mainresult*}

This algorithm and variants of it will be mathematically motivated and discussed in detail in Section \ref{sec:bestheuristic}. The algorithm in Figure \ref{fig:mainresult} is quenched, since it uses the degree sequence. An important variant is the annealed algorithm: for parametric random graph models, one first estimates the degree frequencies, and then the algorithm is applied to the estimated degree sequence. 

For graphs which are not connected, but have a unique giant component, the algorithm can be applied to this giant component. This is the case for sparse Erd\H{o}s-R\'enyi graphs, see Section \ref{sec:sparse}. % In these graphs, the degree distribution in the giant component will depend on the edge density, so that the algorithm is automatically tested for different degree distributions. 

The algorithm also gives the tools to estimate quantities other than the infected fraction, like fluctuations around the metastable equilibrium and correlations between neighboring nodes, see Section \ref{sec:5other}. 

\paragraph{Power and limitations} The algorithm is designed for graphs which are locally tree-like and have a degree distribution with exponential tails. This includes Erd\H{o}s-R\'enyi graphs, random regular graphs and configuration model graphs with suitable degrees. Ultimately, the algorithm relies on the mean field assumption that pairs of nodes with the same pair of degrees have a similar neighborhood which can be well approximated by an `average neighborhood'. This assumption might fail if there are some nodes with very high degree, like in power-law random graphs. Also for grids the algorithm is not expected to work very well, since the grid structure induces much stronger local correlations than a tree. 

The main power of our algorithm is that it incorporates correlations, while still being computationally feasible. In dense graphs, HMF and NIMFA perform quite well. However, it gets more and more delicate to make accurate estimates when graphs get sparser. Sparseness makes averages less reliable, correlations more important and heuristics more sensitive to errors. Our algorithm extends HMF by incorporating correlations. We did extensive simulations (Section \ref{sec:numerics5}), which show that in sparse Erd\H{o}s-R\'enyi graphs the algorithm is much more accurate than HMF and NIMFA (cf. Figure \ref{fig:Heuristics_Simulation}). Higher order methods improve the accuracy, but complicate the model. In practice, a higher order method therefore is often used in an approximate form. The challenge is to reduce the complexity, while still capturing the essential information. We address this problem by a degree-based second order method. Since it is based on pairs of degrees rather than pairs of nodes, the number of variables in the system stays limited.  

The quenched variants of our algorithm typically give more accurate results than the annealed versions, at the cost of more complicated computations. The most advanced quenched version would use the full graph structure and would be an extension of NIMFA. It seems however that the same quality of estimation can already be reached by simpler variants, using only the degree sequence or even only the sum of the degrees (for Erd\H{o}s-R\'enyi graphs this works well).

\paragraph{Computation time} 

Our method introduces for each pair of degrees of neighbors a small system of linear equations. These systems can be solved independently. These equations also depend on a few global parameters, which can be found by an iterative procedure. 

\begin{figure}
	\includegraphics[width=\columnwidth]{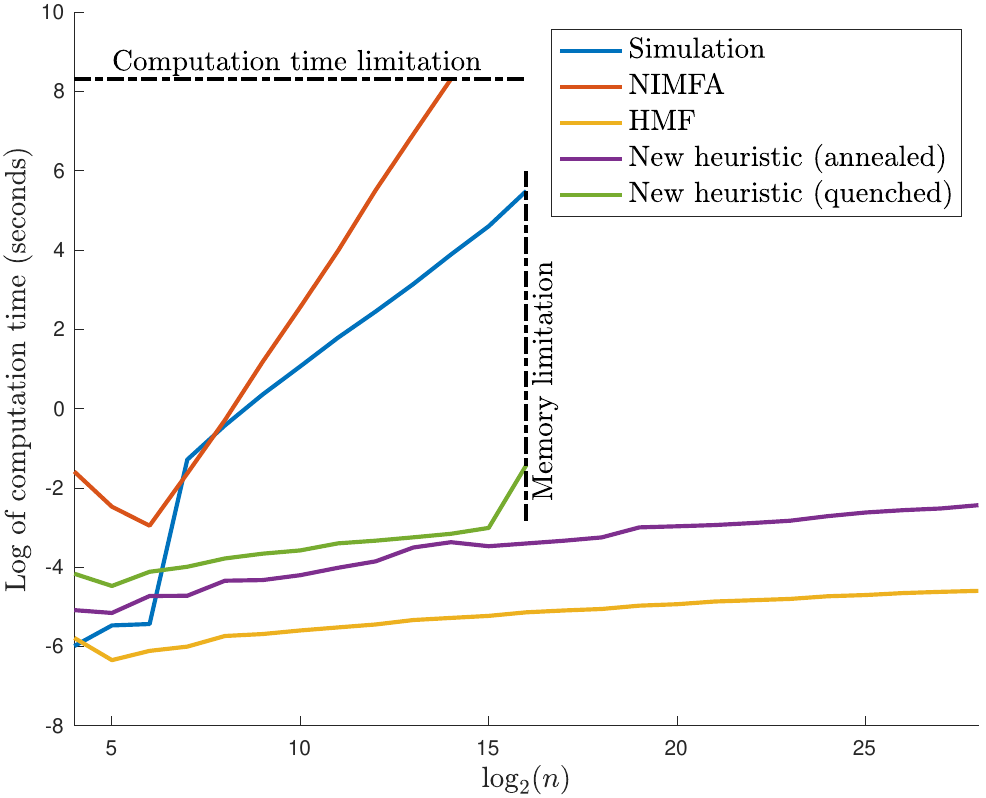}\caption{Comparison of computation times for Erd\H{o}s-R\'enyi graphs with $p=\log(n)/n$ and $\lambda = 2$. NIMFA needs a lot of time. Quenched methods and simulation need a lot of memory.}\label{fig:computime}
\end{figure}

Figure \ref{fig:computime} compares the computation time for different algorithms (the same as in Figure \ref{fig:Heuristics_Simulation}, exact definitions follow later). All computations were done on a normal work laptop. The cheapest algorithms are the annealed algorithms. They require to first compute the binomial probabilities. For degrees greater than twice the expectation, the probability is negligible in this case and approximated by 0. The remaining probabilities are numerically approximated. For HMF, one then has to solve a single non-linear equation. Up to size $n=2^{28}$, the whole computation requires only a fraction of a second. 

The annealed version of our new heuristic is slightly more complicated. Instead of solving one non-linear equation, we have to simultaneously solve about $(2\log(n))^2$ linear systems of four equations with four unknowns and two non-linear equations. Still, the solution is found in less 0.1 seconds for graph sizes up to $n=2^{28}$. The bottleneck for these two annealed methods is the maximal size of numbers the system is able to handle. The computation time itself only grows very slowly with $n$. 

The next method in order of computational complexity is the quenched version of our new heuristic. For this method, we generate a graph and use the actual degree sequence. The most demanding part of the computation is counting to find the degree frequencies (linear in $n$). For graph sizes up to $n=2^{16}$, the algorithm is still completed in less than a second. Bigger graph sizes were out of reach, because memory capacity did not allow for it. For these sparse graphs, the memory size needed grows like $n\cdot\log(n)$. For denser graphs memory space of order $n^2$ might be needed.

The problem of memory also appears when doing simulations. A simulation of $10^3$ time units of the SIS process on $n=2^{16}$ nodes is completed in a few minutes. Memory restrictions are more of an issue than computation time. It has to be noted here that our choice of infection rate allows for a particularly efficient implementation. Other choices for $p$ and $\tau$ might make it harder to run the simulation, both in terms of memory and computation time.

Finally, NIMFA requires to solve a non-linear system of $n$ equations in $n$ unknowns. Now computation time really becomes the deciding factor. Doing the computation for $n=2^{14}$ already takes more than one hour. The methods in \cite{CM12,G11} which do try to incorporate correlations are even much worse from a computational point of view.

\section{Improved heuristics for the infected fraction}\label{sec:improved}

In this section we start to gradually build towards our main estimation method which will be presented in Section \ref{sec:bestheuristic}. We show how to use the degrees to design more accurate estimation methods. One of these methods will turn out to be equivalent to NIMFA, another to HMF. 

The methods in this section, although much better than Heuristic \ref{heuristic1}, still have their shortcomings. We will explain where it goes wrong, and in particular why NIMFA and HMF have a serious bias in sparse graphs. Nevertheless, the ideas in this section are the basis for the exposition of more sophisticated methods in Section \ref{sec:bestheuristic}.

\subsection{Nodes with higher degree are more frequently infected}

Both in the Erd\H{o}s-R\'enyi graph and in the complete graph, the fraction of nodes that is infected is quite stable over time, see again Figure 1. Since nodes are indistinguishable in the complete graph, each individual node is expected to be infected the same fraction of time.  In the complete graph ($n=1000,\lambda = 2$), \emph{all} nodes are infected about half of the time. If we send the running time (not too fast, to avoid extinction) and the number of nodes to infinity, this fraction of time will converge to 1/2 for each individual node. To illustrate, we ran the process on the complete graph for $10^5$ units of time and found that 99\% of the nodes was infected between 49.5 and 50.3\% of the time. 

\begin{figure}
	\includegraphics[width=\columnwidth]{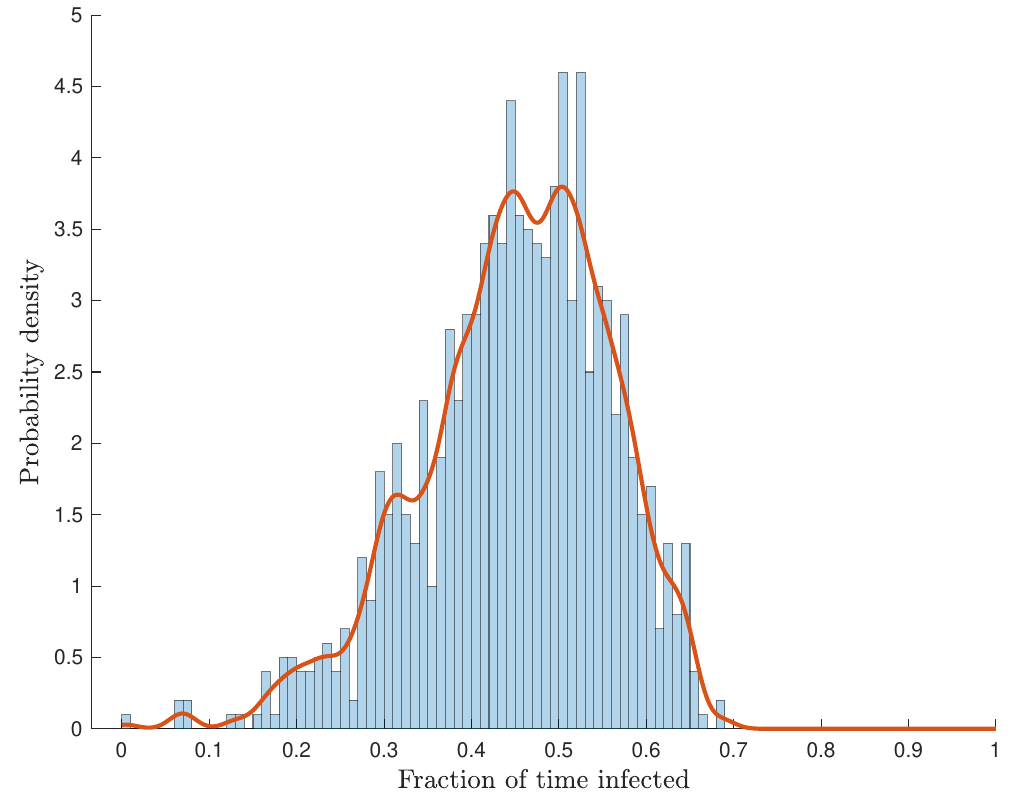}\caption{Simulated and estimated probability density of the fraction of time a node is infected (ER graph with $n=1000$, $p=\log(n)/n$, $\lambda = 2$).}\label{fig:density_inftime}
\end{figure}

The fraction of time an individual node is infected could be quite different in the Erd\H{o}s-R\'enyi graph. In Figure \ref{fig:density_inftime}, we see a histogram of these fractions of time and an estimate of the density function (using kernel density estimation). In this Erd\H{o}s-R\'enyi graph, nodes are only \emph{on average} infected about half of the time. The fraction of time an individual node is infected ranges from 0 to about 0.7. For each individual node it will converge if the running time of the process increases, but the limiting fraction for each node will depend on the geometry of the graph.

The degree of a node can be used to give quite a good estimate for the fraction of time this node will be infected. This in turn will explain the shape of the density function in Figure \ref{fig:density_inftime}. In first approximation, the fraction of time a randomly picked node is infected is equal to $1-\lambda^{-1}$. It therefore infects each of its neighbors at rate 
$
\tau-\tau\lambda^{-1}.%\frac{\tau\mu}{n} = \tau-\frac{1}{np} = \frac{\tau n p-1}{np}.
$

Let $i$ be a random node with degree $d = d_i$. The fraction of time this node is infected will be called $f_i$. Assume that its neighbors are `random' nodes. Then $i$ is healing at rate 1 and, since it has $d$ neighbors, getting infected at rate 
$
d\tau-d\tau\lambda^{-1}.
$

\begin{figure}
	\includegraphics[width=\columnwidth]{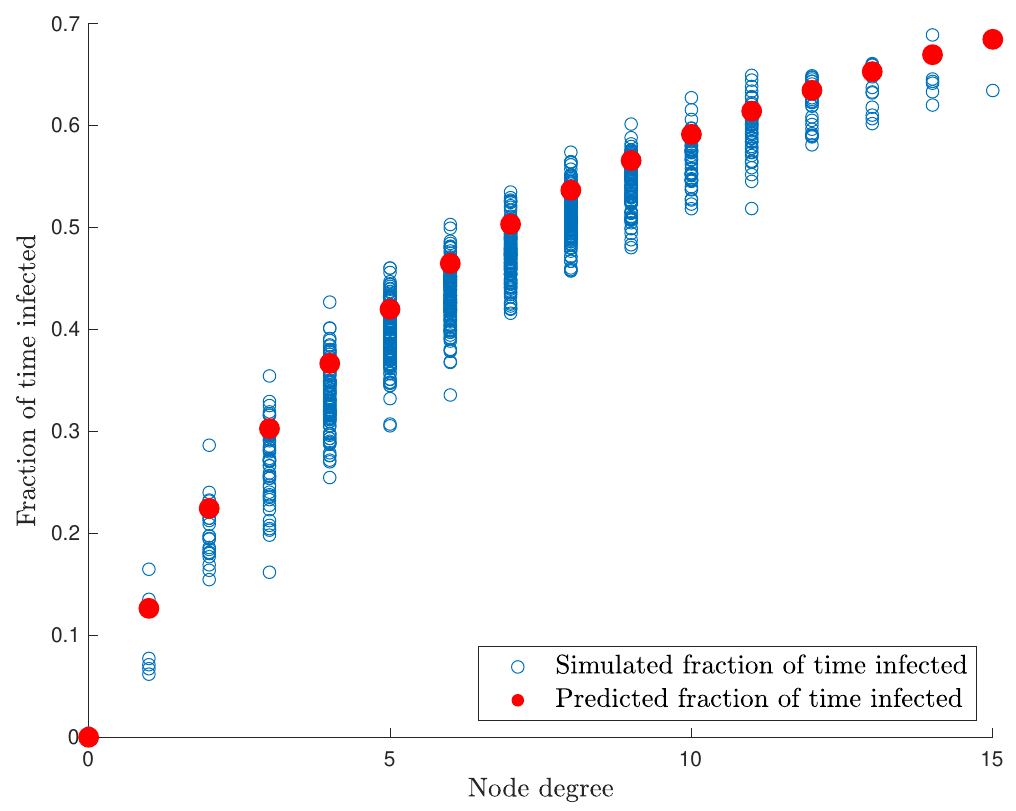}\caption{Simulation of the fraction of time nodes are infected, plotted as points $(d_i,f_i)$ in blue (ER graph with $n=1000$, $p=\log(n)/n$, $\lambda = 2$). The red markers indicate the degree-dependent estimate $(d,f(d))$ based on the heuristic \eqref{eq:degd_prediction}.}\label{fig:fracinf_d}
\end{figure}

If we consider a Markov process with one node, which is healing at rate 1 and getting infected at rate $\alpha$, in the stationary distribution this node is infected a fraction $\alpha/(1+\alpha)$ of time. Therefore, in our more complicated model, the fraction of time $i$ is infected is estimated by 
\begin{align}\label{eq:degd_prediction}
\hat{f}_i := \frac{d\tau-\frac{d\tau}{\lambda}}{1+d\tau-\frac{d\tau}{\lambda}}.% = \frac{d\tau np -d}{np+d\tau n p-d} = \frac{d(\tau - \frac{1}{np})}{1+d(\tau - \frac{1}{np})}.
\end{align}
Fixing the parameters of the graph and the process, this estimate only depends on $d$, so we write $f(d)$ for $\hat{f}_i$. In Figure \ref{fig:fracinf_d}, we plotted the points $(d,f(d))$ and compare with all the points $(d_i,f_i)$ obtained by simulation.

\begin{figure}
	\includegraphics[width=\columnwidth]{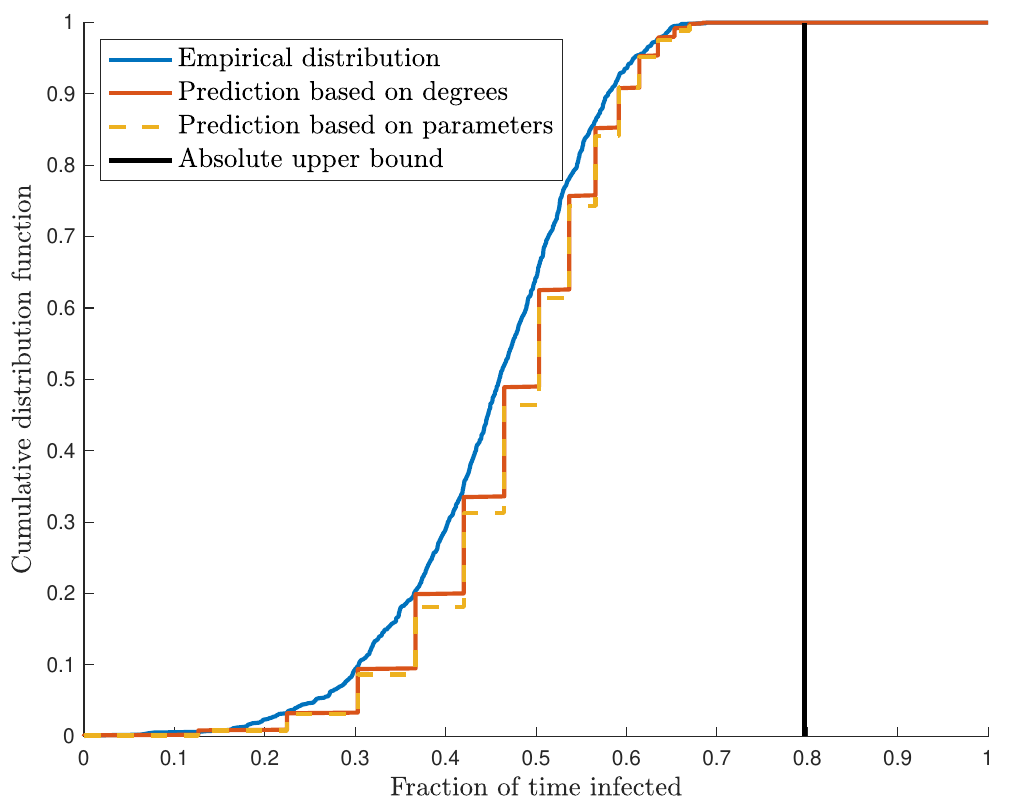}\caption{Cumulative distribution of infected fraction of time (ER graph with $n=1000$, $p=\log(n)/n$, $\lambda = 2$). Blue: empirical. Red: quenched estimate. Yellow: annealed estimate. Both estimates have jumps at the values $f(d)$. Black: absolute upper bound for the infected fraction caused by the maximal degree.}\label{fig:cdf_inftime}
\end{figure}

In Figure \ref{fig:cdf_inftime}, we compare the empirical distribution function of the per-node infected fraction of time,
$
\frac{1}{n}\sum_{i=1}^n \mathbbm{1}\left\{f_i\leq x\right\}, 
$
with the quenched estimated cumulative distribution function 
$
\frac{1}{n}\sum_{i=1}^n \mathbbm{1}\{f(d(i))\leq x\}. 
$

To compute this estimation, we need the actual degrees in the graph. Since we know that the degree distribution $D$ in the graph is $\text{Bin}(n-1,p)$, we can also estimate the distribution without observing the actual degrees by
\begin{align}
\frac{1}{n}\sum_{i=1}^n \mathbbm{1}\{f(d(i))\leq x\} &= \sum_{d=0}^{n-1} \frac{\#\{i:d_i=d\}}{n}\cdot \mathbbm{1}\{f(d)\leq x\}\nonumber\\
&\hspace{-2.5cm}\approx %\sum_{d=0}^{n-1} \mathbb{P}(D=d)\cdot \mathbbm{1}\{f(d)\leq x\}\\
%&= 
\sum_{d=0}^{n-1} \binom{n-1}{d} p^d(1-p)^{n-1-d}\cdot \mathbbm{1}\{f(d)\leq x\}.
\end{align}
% Since the degree is discrete, we obtain a discrete probability distribution. Approximating by
%\[
%\mathbb{P}(D=d) = \mathbb{P}(\text{Bin}(n-1,p) = d)\approx %\mathbb{P}(\text{Pois}(np)=d) = \frac{(np)^d}{d!}e^{-np},
%\]
This estimate as well is compared with the observed empirical distribution in Figure \ref{fig:cdf_inftime}. Note that this is an annealed estimate: it can be computed using only the three parameters $n$, $p$ and $\tau$.

If the maximal degree $\Delta$ in the graph is known, we can compute an absolute upper bound for the (limiting) fraction of time a node is infected. Let $\theta$ be the maximal fraction of time a node is infected. Let $i$ be a node in the graph. If all its neighbors are infected all the time, then $i$ is infected at rate at most $\Delta\tau$. It follows that 
%\begin{align}\label{eq:infupper}
$
\theta \leq \Delta\tau/(1+\Delta\tau) < 1.
$
%\end{align}

A sharper heuristic upper bound is obtained by the using positive correlations between $i$ and its neighbors (shown in \cite{D93,CM14}). The neighbors of $i$ also are infected at most a fraction $\theta$ of time. Positive correlations mean that this upper bound still holds when restricting to periods when $i$ is healthy. Therefore, $i$ is infected at rate at most $\Delta\tau \theta$. This implies that 
%\begin{align}\label{eq:maxinftime}
$
\theta\leq \Delta\tau \theta/(1+\Delta\tau \theta).
$
%\end{align}
It follows that $\theta$ is at most equal to the largest solution of this equation, which is given by
$
(\tau\Delta-1)/(\tau\Delta).
$

In our example in Figure \ref{fig:cdf_inftime}, we have $\tau = 2/(np) = 2/\log(1000)\approx 0.29$ and $\Delta = 17$. This gives approximately $0.79$ as an absolute upper bound for the fraction of time a node can be infected, see the vertical line in the plot.

Note that $\Delta = n-1$ in the complete graph $K_n$, giving the upper bound $1-1/((n-1)\tau)\approx 1-1/\lambda$. This is consistent with known results for $K_n$.

\subsection{Degree-based heuristics for the infected fraction}

The idea of estimate \eqref{eq:degd_prediction} will be used to estimate the infected fraction $\mu/n = \mathbb{E}{\overline{X}}$ of the population. Let $i$ be a node with degree $d$ and assume that its neighbors are infected a fraction $\mu/n$ of time. Analogous to \eqref{eq:degd_prediction}, we estimate the fraction of time $i$ is infected by
\begin{align}\label{eq:fracinf}
\hat f_i := \frac{d\tau \mu/n}{1+d\tau \mu/n}.
\end{align}
A very similar formula is derived in \cite{PV01}, as the stationary solution of the differential equation
\begin{align}\label{eq:HMFdifferential}
\partial_t f(d,t) = -f(d,t) + \tau d (1-f(d,t)) \Theta(t).
\end{align}
Here $f(d,t)$ is the probability a node of degree $d$ is infected at time $t$, and $\Theta(t)$	
is the fraction of half-edges connected to an infected node. Note that $\Theta(t)$ is not exactly the same as the infected fraction of the population at time $t$. We will come back to this in Section \ref{sec:sizebias}. In stationarity, $\Theta(t)$ does not depend on $t$, and setting the derivative in \eqref{eq:HMFdifferential} equal to zero gives the analogue of \eqref{eq:fracinf}.

Averaging the quantity $\hat f_i$ over all nodes in the graph, we obtain an estimate for $\mu/n$. Hence, if $n$, $p$ and $\tau$ are given, $\mu/n$ can be estimated by (numerically) solving
\begin{align}
\frac{\mu}{n} %&= \frac 1 n \sum_{i=1}^n \hat f_i 
%&= \sum_{d=0}^{n-1} \frac{\#\{i:d_i=d\}}{n} \cdot\frac{d\tau \mu/n}{1+d\tau \mu/n}\\
&\approx \sum_{d=0}^{n-1} \binom{n-1}{d}p^d(1-p)^{n-1-d}\cdot\frac{d\tau \mu}{n+d\tau \mu}. \label{eq:fraction_prediction}
\end{align}
In fact the right hand side is the expectation $\mathbb{E}[\tau \mu D/(n+\tau \mu D)]$. Therefore, by Jensen's inequality, \eqref{eq:fraction_prediction} implies 
%	\begin{align}
%		\frac{\mu}{n} \leq \frac{\tau\mu\mathbb{E}[D]}{n+\tau\mu\mathbb{E}[D]}.
%	\end{align}
$
\mu/n \leq \tau\mu\mathbb{E}[D]/(n+\tau\mu\mathbb{E}[D]).
$
Solving for $\mu/n$ gives $\mu/n\leq 1-\lambda^{-1}$, so that this procedure to estimate the infected fraction always gives a lower estimate than Heuristic \ref{heuristic1}.

For large $n$ and $d$, the binomial probabilities in \eqref{eq:fraction_prediction} are hard to compute, but they are well approximated by using the central limit theorem.%, 
%	\begin{align}
%		\mathbb{P}(\text{Bin}(n-&1,p)=d)\approx \mathbb{P}\left(|N((n-1)p,(n-1)p(1-p)) - d|\leq \tfrac12\right) \\
%		%&= \mathbb{P}\left(\left|N(0,1)+\frac{(n-1)p-d}{\sqrt{(n-1)p(1-p)}}\leq \frac{1}{2\sqrt{(n-1)p(1-p)}}\right|\right)\\
%		&= \Phi\left(\frac{d-(n-1)p+\tfrac12}{\sqrt{(n-1)p(1-p)}}\right)-\Phi\left(\frac{d-(n-1)p-\tfrac12}{\sqrt{(n-1)p(1-p)}}\right), \label{eq:normapprox}
%	\end{align} 
%	where $\Phi$ is the cumulative distribution function of the standard normal. 
This gives us 
\begin{heuristic}\label{heuristic2}
	Consider the SIS process on an Erd\H{o}s-R\'enyi graph $G = G_{n,p}$ with infection rate $\tau$. The quasi-stationary infected fraction $\mu/n$ satisfies
	\begin{align}\label{eq:heuristic2}
	\frac{\mu}{n} = \sum_{d=0}^{n-1} P(n,p,d)\cdot\frac{d\tau \mu}{n+d\tau \mu}.
	\end{align}
	Here $P(n,p,d)$ is either
	\begin{enumerate}[(a)]
		\item\label{heuristic2a} The probability $\mathbb{P}(\text{Bin}(n-1,p)=d)$ or an approximation to it (as in \eqref{eq:fraction_prediction}).
		\item\label{heuristic2b} The exact frequency $\#\{i:d_i=d\}/n$.
	\end{enumerate}
\end{heuristic}
Note that \eqref{heuristic2a} gives an annealed estimate, while \eqref{heuristic2b} is a quenched estimate. In view of earlier discussions, we expect  \eqref{heuristic2b} to be more accurate.

For $n\to\infty$, the probability mass of the degree distribution concentrates around its expectation $np$ with standard deviation $\sqrt{np(1-p)}$. This means that for $np\to\infty$ and $\tau np=\lambda$ the right hand side of \eqref{eq:heuristic2} converges to
\[
\lim_{n\to\infty} \frac{np\tau\mu}{n+np\tau\mu} =  \frac{\lambda \lim_{n\to\infty} \mu/n}{1+\lambda\lim_{n\to\infty} \mu/n}. 
\]
Solving \eqref{eq:heuristic2} for $\lim_{n\to\infty}\mu/n$, we obtain 
\begin{align}\label{eq:heur2nlarge}
\lim_{n\to\infty} \frac{\mu}{n} = 1-\frac 1\lambda,
\end{align}
so that both versions of Heuristic \ref{heuristic2} coincide with Heuristic \ref{heuristic1} for $n\to\infty$.

%	\subsubsection{Simulation for $p= 1/\sqrt{n}$}
%	
%	We test Heuristic \ref{heuristic2}\ref{heuristic2a}, taking edge probability $p = 1/\sqrt{n}$ and using the normal approximation \eqref{eq:normapprox}. The average degree in the graph is then approximately $np = \sqrt{n}$. For each $n = 10^2, 11^2,\ldots,50^2$, we generated 25 replications of the Erd\H{o}s-R\'enyi graph. In total we have $1025$ graphs varying in size from 100 to 2500 nodes. On each of those graphs, we simulate the SIS process with infection rate $\tau = 2/\sqrt{n}$. With these choices, the global infection rate is $\lambda = 2$, so the crude Heuristic \ref{heuristic1} predicts that on average half of the population is infected. Our simulation results are given in Figure \ref{fig:heur2sim}. Each simulated graph is represented by a small blue circle. The horizontal coordinate is the expected average degree $\sqrt{n}$. The vertical coordinate is the simulated infected fraction $\overline X$, averaged over time. This means it is an estimate for the quasi-stationary mean. For each value of $n$, the blue line gives the average over the 25 replications. The red horizontal line is Heuristic \ref{heuristic1} and the other red curve is Heuristic \ref{heuristic2}\ref{heuristic2a}. Note that the quasi-stationary mean depends on realization of the graph, but these two heuristics do not. Predictions are the same if graphs have the same horizontal coordinate.

\subsubsection{Simulation and annealed estimation for $p = \log(n)/n$}

We test Heuristic \ref{heuristic2}\ref{heuristic2a}, taking edge probability $p = \log(n)/n$ and using the normal approximation. For each $n=\lceil e^{k/10}\rceil$, $k=45,46,\ldots,80$, we generate 100 replications of the Erd\H{o}s-R\'enyi graph so that in total we have $3600$ graphs. The average degrees vary from $4.5$ to 8 and the sizes range from 91 to 2981 nodes. On each of those graphs, we simulate the SIS process with infection rate $\tau = 2/\log(n)$.

\begin{figure*}
	\includegraphics[width=\textwidth]{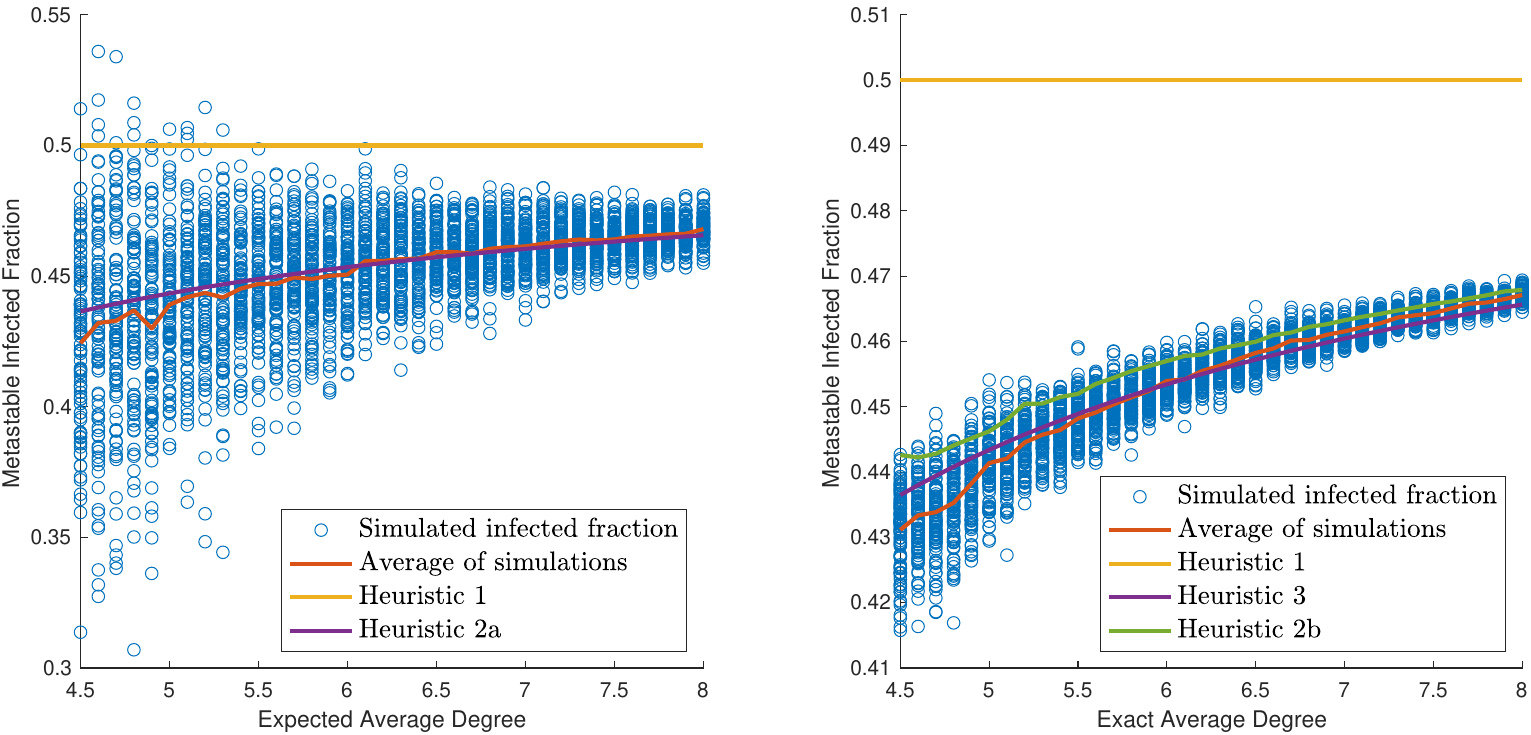}\caption{Infected fractions in Erd\H{o}s-R\'enyi graphs with $p= \log(n)/n$ and $\lambda = 2$. Left: annealed estimate. Right: quenched estimate (the horizontal coordinate of each point depends on the graph.)
	}\label{fig:heur2sim}
\end{figure*}

Our simulation results are given in Figure \ref{fig:heur2sim}, left plot. Each simulated graph is represented by a small blue circle. The horizontal coordinate is the expected average degree $\log(n)$. The vertical coordinate is the simulated infected fraction $\overline X$, averaged over time. This is an estimate for the quasi-stationary mean. The variance is decreasing, since the graph sizes are increasing from left to right.
%		\begin{figure}
%		\begin{center}
%			\includegraphics[scale=0.45]{heur2sim2.pdf}\caption{The annealed Heuristic \ref{heuristic2}\ref{heuristic2a} compared to simulation for $p= \log(n)/n$ and $\tau = 2/\log(n)$. Horizontal axis: expected average degree. Vertical axis: quasi-stationary infected fraction of the population.}\label{fig:heur2sim}
%		\end{center}
%	\end{figure}

\subsubsection{Using graph information: quenched estimates}

Heuristic \ref{heuristic2}\ref{heuristic2a}  seems to do quite well on average in Figure \ref{fig:heur2sim}, but there is a lot of variation in the simulation results. 
The estimate by Heuristic \ref{heuristic2}\ref{heuristic2a} can be made \emph{before} generating the graph. Can we estimate the stationary infected fraction more accurately \emph{after} generating the graph and using characteristics of the graph?

%	\begin{figure}
%		\begin{center}
%		\includegraphics[scale=0.45]{heur2simfixed3.pdf}\caption{The quenched Heuristics \ref{heuristic2}\ref{heuristic2b} (green) and \ref{heuristic3} (red) compared to simulation (blue) for $p= \log(n)/n$ and $\tau = 2/\log(n)$. The horizontal axis gives the \emph{exact} average degree.}\label{fig:heur2simfixed}
%		\end{center}
%	\end{figure}

To test this, we again generated Erd\H{o}s-R\'enyi graphs with the same numbers of nodes as in Figure \ref{fig:heur2sim}. If $p=\log(n)/n$, the expected total number of edges is $p\binom{n}{2} = \tfrac12 (n-1)\log(n)$. This time we conditioned the graphs to have the number of edges exactly equal to its (rounded) expectation. We can even eliminate the rounding error by slightly adjusting $p$. The simulation results are shown in Figure \ref{fig:heur2sim}, right plot. The horizontal coordinate now is the \emph{exact} average degree, since we fixed the number of edges to its expectation. For unconditioned ER graphs, this exact average degree is a quenched parameter, which can only be computed after observing the graph. 

It turns out that graphs with the same exact average degree have a lot less variation in their quasi-stationary means than graphs merely having the same expected average degree. A lot of the variation is explained by variation of the number of edges in the graph. The estimate therefore becomes more accurate by the following quenched estimation method: first estimate $p$ based on the observed number of edges, then use this adjusted $p$ for estimating the metastable infected fraction. The resulting heuristic is:

%\addtocounter{heuristic}{-1}
\begin{heuristic}\label{heuristic3}
	Let $G$ be an Erd\H{o}s-R\'enyi graph with $n$ nodes and $m$ edges.
	Consider the SIS process on $G$ with infection rate $\tau$. The quasi-stationary infected fraction $\mu/n$ satisfies
	\begin{align}\label{eq:heuristic3}
	\frac{\mu}{n} = \sum_{d=0}^{n-1} P\left(n,\frac{m}{\binom{n}{2}},d\right)\cdot\frac{d\tau \mu}{n+d\tau \mu}.
	\end{align}
	Here $P(n,p,d)$ is (an approximation to) the probability $\mathbb{P}(\text{Bin}(n-1,p)=d)$.
\end{heuristic}
For the figures, we use the normal approximation. This estimate still seems to have a small systematic bias for some values of $n$. We will discuss systematic errors in Heuristic \ref{heuristic2} and \ref{heuristic3} in the next section.

Heuristic \ref{heuristic3} uses the exact number of edges in the graph, which is equivalent to using the sum of the degrees. Next steps would be to use all individual degrees or even the full adjacency matrix $A$ of the graph. If the full degree sequence is known, we could use Heuristic \ref{heuristic2}\ref{heuristic2b} with the exact numbers $\#\{i:d_i=d\}$. It turns out by simulations that this is more accurate than Heuristic \ref{heuristic2}\ref{heuristic2a}, but has a larger bias than Heuristic \ref{heuristic3} in the simulated range, see again Figure \ref{fig:heur2sim}. A remark to be made here is that now graphs with the same exact average degree might get different estimates. The curve for Heuristic \ref{heuristic2}\ref{heuristic2b} only shows the average of these estimates. This means that estimation errors for individual graph realizations can not be seen in the picture.

\begin{figure}
	\includegraphics[width=\columnwidth]{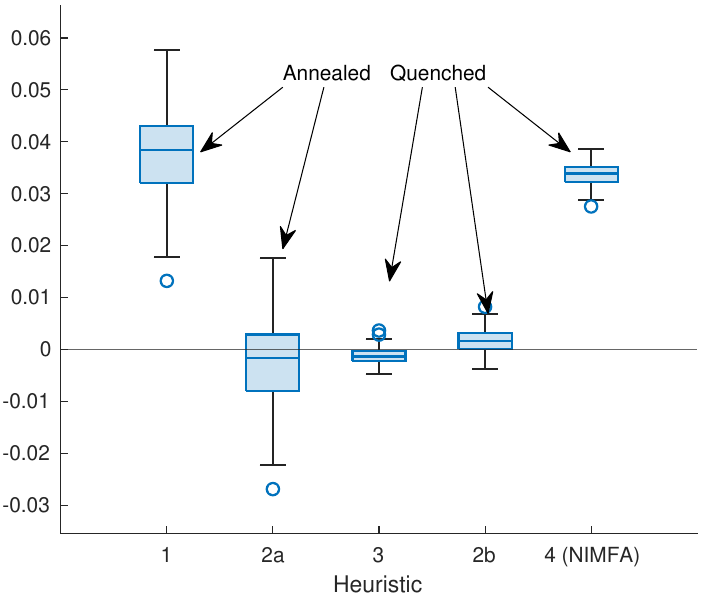}\caption{Error boxplots for each heuristic for Erd\H{o}s-R\'enyi graphs with $n=1000$, $p = \log(n)/n$ and $\lambda = 2$. The errors of annealed methods have a standard deviation of about 0.01, cf. Figure \ref{fig:annealedquenched}. The errors of quenched heuristics have smaller variation. Heuristic \ref{heuristic1} and NIMFA are clearly biased.}\label{fig:heuristics_boxplots}
\end{figure}

To see individual errors, we simulated the SIS process on 100 Erd\H{o}s-R\'enyi graphs with $n=1000$ nodes, $p=\log(n)/n$ and $\lambda = 2$. Figure \ref{fig:heuristics_boxplots} gives boxplots for the errors: the difference between the estimated infected fraction and the simulated infected fraction for the different heuristics. The errors for the annealed heuristics have much more variation. On the other hand, more detailed information than the number of edges does not really improve the estimation: Heuristic \ref{heuristic3} is not worse than Heuristic \ref{heuristic2}\ref{heuristic2b} and \ref{heuristic4} (these heuristics respectively use the exact number of edges, the exact degrees and the full adjacency matrix). Heuristic 4 (to be discussed next) has a small variation, but a large systematic error. The systematic errors of Heuristics \ref{heuristic3} and \ref{heuristic2}\ref{heuristic2b} are relatively small, but the situation gets worse when the graph is more sparse, in particular when the graph disconnects. In that case one has to be more careful, see Section \ref{sec:numerics5}. In general, these heuristics work quite well on graphs which are fairly homogeneous such as the complete graph or Erd\H{o}s-R\'enyi graph with enough edges. For less homogeneous graphs like power-law graphs, other heuristics are needed.

Now assume the full adjacency matrix is known. Writing again $f_i$ for the fraction of time $i$ is infected, and assuming independence between nodes, we find that $i$ gets infected at rate $\tau\sum_{j:i\sim j} f_j$. Therefore $f_i$ satisfies
\begin{align}\label{eq:infect_system}
f_i = \frac{\tau\sum_{j:i\sim j} f_j}{1+\tau\sum_{j:i\sim j} f_j}.
\end{align} 
It turns out that this is equivalent to NIMFA, which is derived as follows. Consider $\mathbb{P}(X_i(t)=1)=\mathbb{E}[X_i(t)]$. By standard Markov chain theory, its derivative is expressed in the transition rates by
\begin{align*}
\frac{d}{dt}\mathbb{E}[X_i(t)] = -\mathbb{E}[X_i(t)] + \tau \sum_{j:i\sim j} \mathbb{E}[(1-X_i(t))\cdot X_j(t)],
\end{align*}
where the first term on the right corresponds to healing of $i$ and the second to infection.
The NIMFA assumption is to ignore correlations by setting 
\begin{align}\mathbb{E}[X_i(t)\cdot X_j(t)] = \mathbb{E}[X_i(t)]\cdot\mathbb{E}[X_j(t)].\end{align}
The stationary solution is then obtained by solving
\begin{align}
0 = -\mathbb{E}[X_i]+\tau \sum_{j:i\sim j} \mathbb{E}[X_j] + \tau\mathbb{E}[X_i] \sum_{j:i\sim j} \mathbb{E}[X_j],
\end{align}
which is the same as \eqref{eq:infect_system}. Solving this system of $n$ non-linear equations and $n$ unknowns gives a quenched estimate $\hat f_i$, after which the infected fraction of the population can be estimated by 
\begin{heuristic}[NIMFA]\label{heuristic4}
	Let $G=(V,E)$ be an Erd\H{o}s-R\'enyi graph. Consider the SIS process on $G$ with infection rate $\tau$. Let $\hat f_i,i\in V$ be the solution of the system \eqref{eq:infect_system}. The quasi-stationary infected fraction $\mu/n$ satisfies
	\begin{align}
	\frac{\mu}{n} = \frac 1 n \sum_{i=1}^n \hat f_i.
	\end{align}
\end{heuristic}
One would expect this to be more accurate than Heuristic \ref{heuristic2}\ref{heuristic2b}. However, our simulations show NIMFA to be strikingly more biased for $n=1000$ and $p=\log(n)/n$ (see Figure \ref{fig:heuristics_boxplots}). This is a general problem with NIMFA in graphs with small degrees. In the next sections we explain how this is possible.

\subsection{Systematic errors I: size bias and the infection paradox}\label{sec:sizebias}

Heuristics \ref{heuristic2}\ref{heuristic2a}, \ref{heuristic2}\ref{heuristic2b} and \ref{heuristic3} are all based on the assumption that the neighbors of a random node $i$ are infected a fraction $\mu/n$ of time. However, given the information that $j$ is a neighbor of $i$, the degree distribution of $j$ is different. This is the so-called friendship paradox, and the degree distribution of $j$ is called the excess degree distribution or size biased distribution \cite{N18,Hofstad17}). Precisely, the probability that $j$ has degree $d$ has to be weighted by a factor $d$. When all nodes independently have the same degree distribution $D$, the degree $D_j$ of $j$ satisfies
\begin{align}\label{eq:sizebias}
\mathbb{P}(D_j=d\mid i\sim j) = %\frac{d\cdot\mathbb{P}(D=d)}{\sum_{d}d\cdot\mathbb{P}(D=d)} =
\frac{d\cdot\mathbb{P}(D=d)}{\mathbb{E}[D]}.
\end{align} 
For the Erd\H{o}s-R\'enyi graph, an exact expression is
\begin{align}
\mathbb{P}(D_j = d\mid i\sim j) %&= \frac{\mathbb{P}(d(u)=d,u\leftrightarrow v)}{\mathbb{P}(u\leftrightarrow v)}\\
%& = \frac{p\cdot \mathbb{P}(\text{Bin}(n-2,p)=d-1)}{p}
%&= \mathbb{P}(\text{Bin}(n-2,p)=d-1) 
= \binom{n-2}{d-1}p^{d-1}(1-p)^{n-d-1}.
\end{align}
For large $n$ and small $p$ and $d$, this results in 
\begin{align}
\mathbb{P}(D_j = d\mid i\sim j) \approx \frac{(np)^{d-1}(1-p)^n}{(d-1)!}	
\approx \frac{d\cdot \mathbb{P}(D_i = d)}{np},
\end{align}
which is consistent with the general formula \eqref{eq:sizebias}. Note that the formula in the middle is also approximately $\mathbb{P}(D_i = d-1)$, so that the degree distribution of $j$ is obtained by shifting the original degree distribution by 1. Hence, in Heuristics \ref{heuristic2}\ref{heuristic2a}, \ref{heuristic2}\ref{heuristic2b} and \ref{heuristic3} we make an error of 1 in counting the neighbors of $j$. This error causes an underestimation of the fraction of time neighbors of $i$ are infected, which in turn also leads to underestimation for $i$ itself. 

\begin{figure}
	\includegraphics[width=\columnwidth]{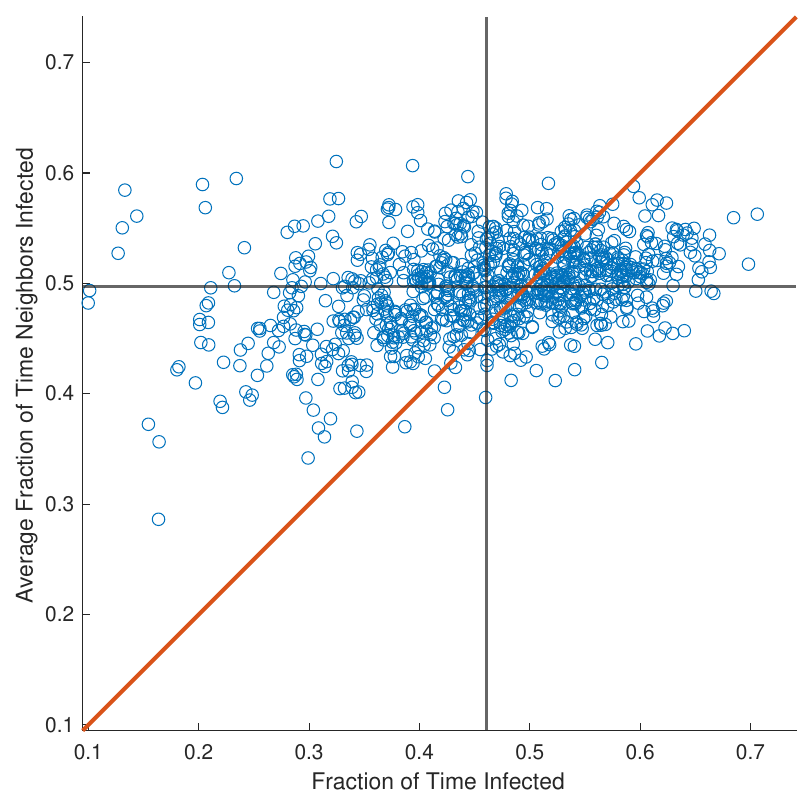}\caption{The infection paradox (ER graph with $n=1000$, $p=\log(n)/n$, $\lambda = 2$): most nodes are less often infected than their neighbors. Averages are indicated by the horizontal and vertical line, they intersect clearly above the diagonal.}\label{fig:infectionparadox}
\end{figure}

The friendship paradox thus implies an infection paradox (illustrated in Figure \ref{fig:infectionparadox}):
\begin{paradox}
	An average neighbor of an average node is more often infected than the node itself. 	
\end{paradox} 

Instead of assuming that $i$ and $j$ themselves have the same degree distribution, it is better assume that neighbors of $j$ have the same degree distribution as neighbors of $i$. Note that this assumption is still not entirely correct, we now essentially make an error of 1 in counting neighbors at distance 2 of $j$. In a locally tree-like graph this error is of a smaller order. Under this improved assumption, we obtain
\begin{align}\label{eq:mutilde0}
\frac{\tilde \mu}{n} = \sum_{d=1}^{n-1} \mathbb{P}(\text{Bin}(n-2,p)=d-1))\cdot \frac{d\tau\tilde\mu}{n+d\tau\tilde\mu}
\end{align}
where $\tilde \mu/n$ is the fraction of time a random neighbor of a random node is infected. This motivates to adapt Heuristic \ref{heuristic2} and to estimate $\mu/n$ by
\addtocounter{heuristic}{-3}
\begin{heuristic}[Variant HMF]  Solve \eqref{eq:mutilde0} for $\tilde\mu$. Then $\mu$ satisfies
	\begin{align}\label{eq:mu0}
	\frac{\mu}{n} = \sum_{d=0}^{n-1} \mathbb{P}(\emph{\text{Bin}}(n-1,p)=d))\cdot \frac{d\tau\tilde\mu}{n+d\tau\tilde\mu}.
	\end{align}
\end{heuristic}
\addtocounter{heuristic}{2}
In fact, $\tilde \mu/n$ is the same as the stationary solution for $\Theta$ appearing in \eqref{eq:HMFdifferential}. Moreover, though the derivation is slightly different, our formula \eqref{eq:mu0} is exactly the same as the Heterogenous Mean Field method designed by Pastor-Satorras and Vespignani in \cite{PV01}. HMF in the form above is an annealed method. It can easily be turned into a quenched method by using the actual degree frequencies.  

The infection paradox leads to underestimation of $\mu/n$, so the adaptation in \eqref{eq:mu0} will increase our previous estimates (Heuristics \ref{heuristic2}\ref{heuristic2a}, \ref{heuristic2}\ref{heuristic2b} and \ref{heuristic3}) for the infected fraction. The same is true for the quenched version of HMF.

\subsubsection{Mean-field methods NIMFA and HMF and size bias}

\begin{figure}
	\includegraphics[width=\columnwidth]{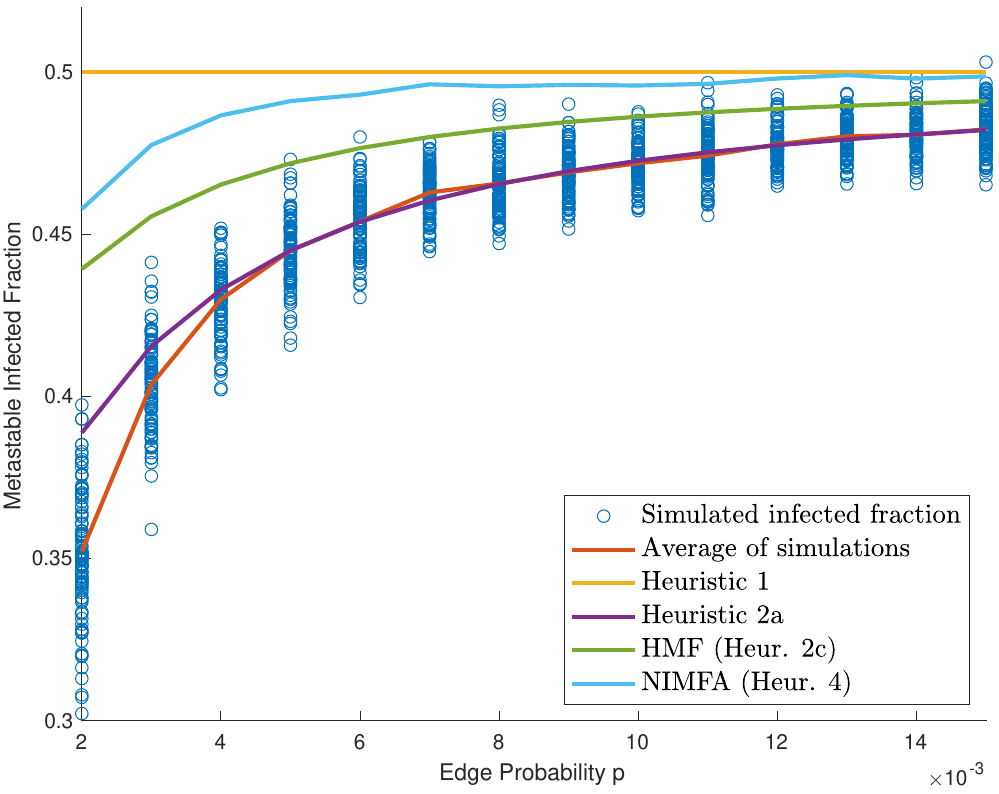}\caption{NIMFA and HMF systematically overestimate the infected fraction and are outperformed by Heuristic \ref{heuristic2}\ref{heuristic2a}. (ER graph, $n=1000$, $\lambda = 2$)}\label{fig:MFbias}
\end{figure}

NIMFA uses the full adjacency matrix, which means that degrees of neighbors are always counted correctly. NIMFA therefore does not make a systematic error caused by size bias. The same is true for HMF, which explicitly takes the size bias effect into account. This explains why both these methods give higher estimates for the infected fraction than for instance Heuristic \ref{heuristic2}\ref{heuristic2a}. This can be clearly seen in Figure \ref{fig:MFbias}. We simulated ER graphs with varying edge probability. For each $p$, 100 graphs were generated, and on each of them we simulated the SIS process. Again we see that Heuristic \ref{heuristic1} is too simple, but \ref{heuristic2}\ref{heuristic2a} is much more accurate than HMF and NIMFA. Only when the graphs become really sparse with average degrees less than 5, Heuristic \ref{heuristic2}\ref{heuristic2a} visibly deviates from the simulation results. We will come back to this in Section \ref{sec:numerics5}.

By correcting for size bias, NIMFA and HMF actually become \emph{more} biased than other methods. The reason is that there are two sources of systematic errors, which work in opposite directions and which often seem to almost cancel each other. The second error source is caused by correlations, and will be discussed in the next section. By repairing only one of these errors, NIMFA and HMF get their bias, which especially will be visible in sparse graphs.

\subsection{Systematic errors II: neighbor correlation}

So far we ignored dependence between neighboring nodes. For NIMFA, this is an explicit assumption. Also the other heuristics discussed before do not take dependence into account. Neighbors tend to align into the same state. When a node $i$ is healthy, this increases the likelihood of its neighbors being healthy as well, see \cite{D93,CM14}. By ignoring this, we overestimate the rate at which $i$ gets infected by its neighbors. For this reason, the paper introducing NIMFA \cite{MOK09} already mentions that the method gives an upper bound for the infected fraction. 

\begin{figure}
	\includegraphics[width=\columnwidth]{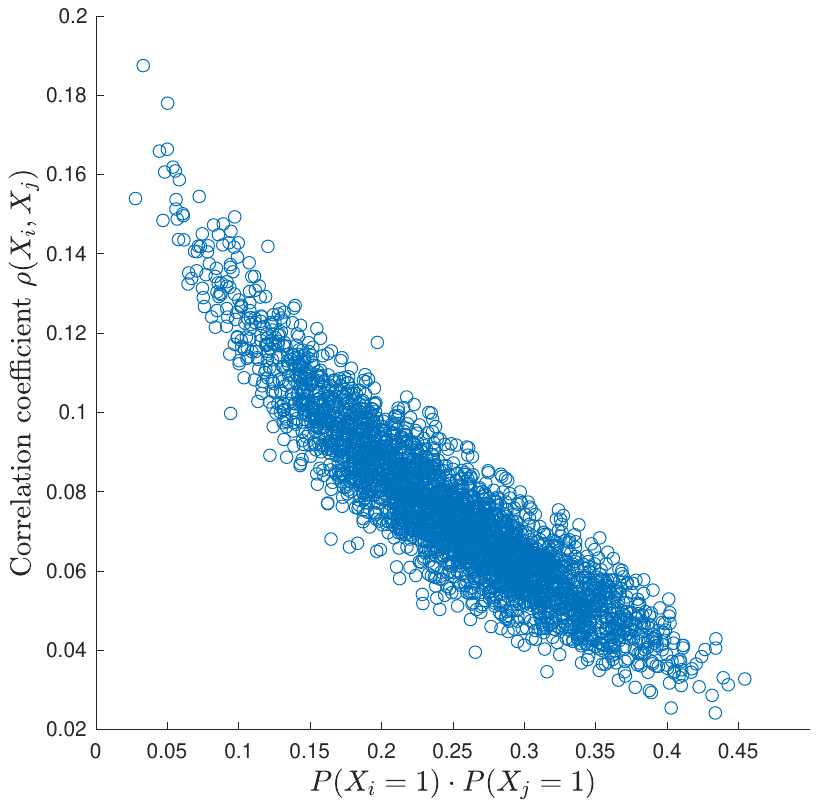}\caption{Neighbor correlations in Erd\H{o}s-R\'enyi graph ($n=1000$, $p = \log(n)/n$, $\lambda = 2$). Each marker represents an edge $(i,j)$. }\label{fig:correlation}
\end{figure}

We illustrate the phenomenon of neighbor correlation in Figure \ref{fig:correlation}. For each edge $(i,j)$ in the graph, (co)variances are given by
\begin{align}\label{eq:covar}
%\text{Var}(X_i) &= \mathbb{E}(X_i^2)-(\mathbb{E}(X_i))^2 = \mathbb{P}(X_i=1)-\mathbb{P}(X_i=1)^2,\\
\text{Cov}(X_i,X_j) &= \mathbb{E}(X_iX_j)-\mathbb{E}X_i\cdot\mathbb{E}X_j\nonumber\\
&\hspace{-1.5cm} =\mathbb{P}(X_i=X_j=1)-\mathbb{P}(X_i=1)\cdot\mathbb{P}(X_j=1).
\end{align}
These are estimated by the corresponding simulated fractions of time, after which we can also estimate the correlation coefficients
\begin{align}\label{eq:correlation}
\rho(X_i,X_j) = \frac{\text{Cov}(X_i,X_j)}{\sqrt{\text{Var}(X_i) \text{Var}(X_j) }}.
\end{align} 
The figure plots the estimated correlation coefficients against the estimated product $\mathbb{P}(X_i=1)\cdot \mathbb{P}(X_j=1)$. Note that the correlation is weaker if the nodes are more often infected. A node which is frequently infected has more neighbors, so that the influence of an individual neighbor is less important.

\begin{figure*}
	\includegraphics[width=\textwidth]{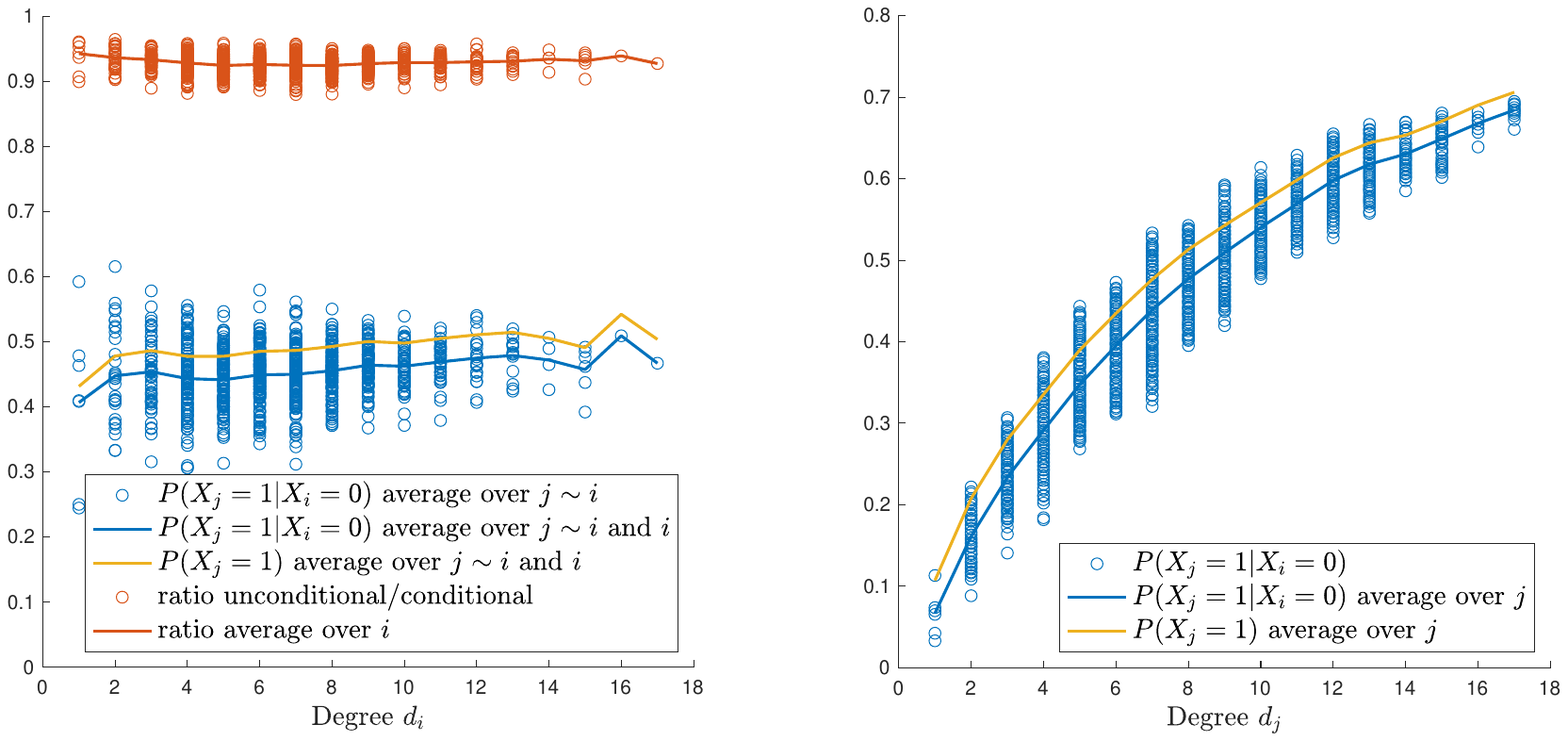}\caption{Dependence of conditional probabilities $\mathbb{P}(X_j=1\mid X_i=0)$ on the degrees $d_i$ and $d_j$.  (ER graph $n=1000$, $p = \log(n)/n$, $\lambda = 2$) }\label{fig:condprob}
\end{figure*}

All heuristics in the previous section use the assumption that the rate at which node $i$ is infected by node $j$ is equal to $\tau\cdot\mathbb{P}(X_j=1)$. However, node $j$ is only able to succesfully infect $i$ if $i$ is healthy. Therefore, the actual rate at which $j$ infects $i$ is equal to $\tau\cdot\mathbb{P}(X_j=1 \mid X_i=0)$, which is smaller than $\tau\cdot\mathbb{P}(X_j=1)$. In Figure \ref{fig:condprob}, we see how the conditional probability $\mathbb{P}(X_j=1 \mid X_i=0)$ depends on the degree of $i$ and the degree of $j$. 

The left plot shows for each node $i$ a simulated estimate of 
\begin{align}\label{eq:condprob}
\frac{1}{d_i}\sum_{j:i\sim j}\mathbb{P}(X_j=1 \mid X_i=0),
\end{align}
the average fraction of time its neighbors are infected, given that $i$ itself is healthy (blue markers). The plot shows that this quantity only mildly depends on the degree of $i$. The variation in this average of conditional probabilities is mainly explained by the variation in the degrees of the \emph{neighbors} of $i$. Taking for each node $i$ the ratio
\begin{align}\label{eq:probratio}
\frac{\sum_{j:i\sim j}\mathbb{P}(X_j=1 \mid X_i=0)}{\sum_{j:i\sim j}\mathbb{P}(X_j=1)},
\end{align}
we get very little variation (red markers). It therefore seems quite reasonable to base heuristics on the assumption that 
\begin{align}
\frac{1}{d_i}\sum_{j:i\sim j}\mathbb{P}(X_j=1 \mid X_i=0) \approx \frac{\eta}{d_i} \sum_{j:i\sim j}\mathbb{P}(X_j=1)
\end{align}
for some constant $0<\eta<1$ which does not depend on $i$ and to try to estimate $\eta$. For each degree $d$, we also plotted the average of the conditional and of the unconditional probabilities, 
\begin{align}
&\frac{1}{\#\{i:d_i=d\}} \sum_{i:d_i=d}\frac 1 d\sum_{j:i\sim j}\mathbb{P}(X_j=1 \mid X_i=0),\\ &\frac{1}{\#\{i:d_i=d\}} \sum_{i:d_i=d}\frac 1 d\sum_{j:i\sim j}\mathbb{P}(X_j=1),
\end{align}
and their ratio's \eqref{eq:probratio} averaged over $i$ (red). From the values on the yellow curve, we can reproduce an estimate of the size biased infected fraction $\tilde \mu/n$ by taking a weighted average:
\begin{align}
\frac{\tilde \mu}{n} = \frac{1}{n} \sum_{d=1}^{n-1}\sum_{i:d_i=d}\frac 1 d\sum_{j:i\sim j}\mathbb{P}(X_j=1).
\end{align}
Dependence of $\mathbb{P}(X_j=1 \mid X_i=0)$ on the degree of $j$ is visible in the right plot of Figure \ref{fig:condprob}. Each blue marker now corresponds to a single ordered edge $(i,j)$. Dependence of $\mathbb{P}(X_j=1)$ on the degree of $j$ has already been observed before, and here we see very similar patterns for the conditional probabilities $\mathbb{P}(X_j=1\mid X_i=0)$. Averages of conditional and of unconditional probabilities are again given by the curves. 

\section{Taking size bias and neighbor correlations into account}\label{sec:bestheuristic}

The two types of systematic errors work in opposite directions. Ignoring the size bias effect leads to underestimation of the infected fraction $\mu/n$ of the population. On the other hand, ignoring correlation leads to overestimation of the infected fraction of the population.

%	It seems that these two opposite errors somehow cancel each other in Heuristics \ref{heuristic2} and \ref{heuristic3}, at least for the parameter values in Figure \ref{fig:heuristics_boxplots}. In Heuristic \ref{heuristic4}, we do not use the binomial degree distribution anymore, but the exact adjacency matrix. This means that Heuristic \ref{heuristic4} does not make the size-bias error. However, it still does make the correlation error. This explains why Heuristic \ref{heuristic4} in the end gives a much larger systematic prediction error, see Figure \ref{fig:heuristics_boxplots}. 

\begin{figure}
	\includegraphics[width = \columnwidth]{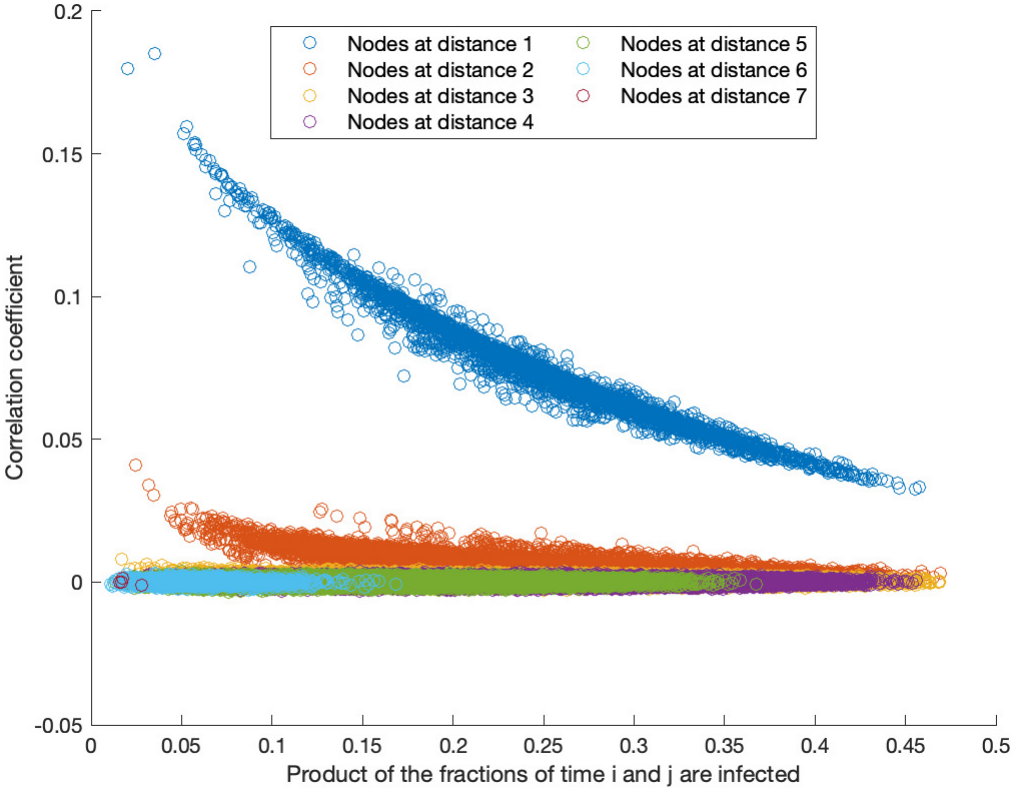}
	\caption{Correlations for each pair $\{i,j\}$ of nodes in an ER graph with $n=1000$, $p=\log(n)/n$ and $\lambda=2$. The color indicates the distance in the graph.}\label{distance_corr}
\end{figure}

The two errors might cancel sometimes, but in general this will not be the case, see for instance the smaller degrees in Figure \ref{fig:MFbias}. We therefore will propose heuristics to repair both errors. Nodes are still positively correlated if they are not direct neighbors. However, correlations rapidly decrease when the distance increases, Figure \ref{distance_corr}. Based on this observation, we will design a heuristic which takes correlations between direct neighbors into account, but ignores correlations between other pairs of nodes. 

Consider random nodes $i$ and $j$ which are neighbors and have degrees $d_i\geq 1$ and  $d_j\geq 1$. We make the following assumptions:
\begin{enumerate}
	%\item $i$ is infected a fraction $\mu_d/n$ of time, (Wordt niet gebruikt!)
	%\item Node $j$ has `degree' $np+1$ (approximate size-biased expectation). Quotations mean that this number can not be a true degree if it is not an integer.
	%\item $j$ is infected a fraction $\tilde \mu_d/n$ of time, (Wordt niet gebruikt!)
	\item All neighbors of $i$ and $j$ (except $i,j$ themselves) are infected the same fraction of time. We assume this fraction to be independent of $d_i$ and $d_j$, and equal to the (unknown) size biased infected fraction $\tilde\mu/n$:
	\[ 
	\mathbb{P}(X_k=1)=\tilde\mu/n,\ \text{for}\ k\sim i\ \text{or}\ k\sim j, k\neq i,j.
	\] 
	\item All these neighbors are assumed to be independent of each other.
	\item There exists a constant $\eta\in (0,1)$ such that 
	%		\begin{itemize}
	%			\item $\mathbb{P}(X_k=1\mid X_i = 0) = \eta\tilde\mu/n$ for 
	%			each $k\sim i, k\neq j$.
	%			\item $\mathbb{P}(X_k=1\mid X_j = 0) = \eta\tilde\mu/n$ for each $k\sim j, k\neq i$.
	%		\end{itemize}
	\[
	\mathbb{P}(X_k=1\mid X_i = 0) = \eta\tilde\mu/n,\ \text{for}\ k\sim i, k\neq j,
	\]
	and similarly for $i$ and $j$ reversed.
\end{enumerate} 
Note that we do not use degrees other than the degrees of $i$ and $j$. This causes the first assumption to be quite inaccurate for individual nodes, but it does keep the analysis feasible and the errors will mostly average out. 

\begin{figure}
	\begin{tikzpicture}[->,>=stealth,scale=.75,auto,bend left]
	%\useasboundingbox (-1,-0.5) rectangle (3,1.8);
	\node[circle,draw,inner sep=-2pt,minimum width=5mm] (1) {\begin{tabular}{c}(1)\\$X_i=0$\\$X_j=0$\end{tabular}};
	\node[circle,draw,inner sep=-2pt,minimum width=5mm] (2) at ($ (1) +  (-30:5) $) {\begin{tabular}{c}(3)\\$X_i=1$\\$X_j=0$\end{tabular}};
	\node[circle,draw,inner sep=-2pt,minimum width=5mm] (3) at ($ (2) +  (30:5)$) {\begin{tabular}{c}(4)\\$X_i=1$\\$X_j=1$\end{tabular}};
	\node[circle,draw,inner sep=-2pt,minimum width=5mm] (4) at ($ (3) + (150:5)$) {\begin{tabular}{c}(2)\\$X_i=0$\\$X_j=1$\end{tabular}};
	
	\path (1) edge[bend right] node[below left]{$\displaystyle\frac{(d_i-1)\eta\tilde\mu\tau}{n}$} (2)
	(1) edge node{$\displaystyle\frac{(d_j-1)\eta\tilde\mu\tau}{n}$} (4)
	(2) edge[bend right] node[below right]{$\displaystyle\frac{(d_j-1)\eta\tilde\mu\tau}{n}+\tau$} (3)
	(2) edge[bend right] node[below left]{1} (1)
	(3) edge node[above right]{1} (4)
	(3) edge[bend right] node{1} (2)
	(4) edge node[above left]{1} (1)
	(4) edge node[above right]{$\displaystyle\frac{(d_i-1)\eta\tilde\mu\tau}{n}+\tau$} (3);
	\end{tikzpicture}
	%\vspace{-.5cm}
	\caption{Transition diagram for infectious status of $i$ and $j$. States are numbered $(1),\ldots,(4)$.}
	\label{fig:transitiondiagram}		
\end{figure}
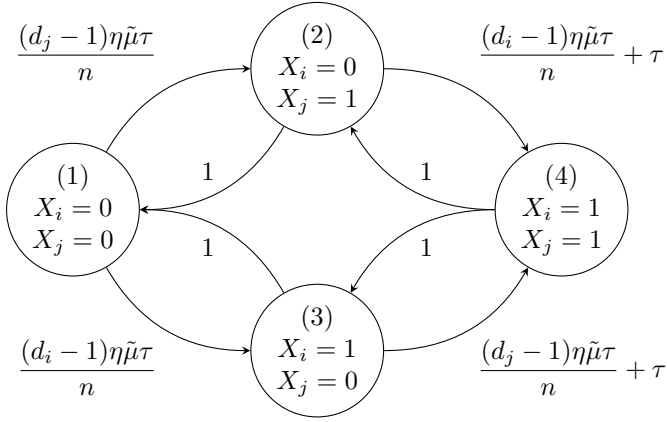

Under these assumptions, we are interested in the simultaneous evolution of nodes $i$ and $j$. This evolution is described by a Markov chain on four states, with transition rates in the diagram in Figure \ref{fig:transitiondiagram}.

This Markov chain has a unique stationary distribution, which is found by solving the equations \eqref{eq:corsystem} on page~\pageref{fig:mainresult}, where $Q$ as given in \eqref{eq:Qmatrix} is the generator matrix.  
%	\begin{align}
%		\begin{cases}\mathbf{x}^TQ =\mathbf{0}^T,\\ \mathbf{x}^T\mathbf{1} = 1,\end{cases} \qquad
%		\mathbf{x}^T = \begin{pmatrix}x_1 & x_2& x_3& x_4\end{pmatrix},
%	\end{align}
%	where $\mathbf{0}$ and $\mathbf{1}$ are the all zero and all one vector in $\mathbb{R}^4$ and $Q=(Q_{k,\ell})$ is the generator matrix:
%	\begin{align}
%		%Q = \begin{pmatrix}-Q_{1,2}-Q_{1,3}&p\eta\tilde\mu\tau&(d-1)\eta\tilde\mu\tau/n&0\\
%		%1 & -Q_{2,1}-Q_{2,4} & 0 & ((d-1)\eta\tilde\mu\tau/n)+\tau\\
%		%1 & 0 & -Q_{3,1}-Q_{3,4} & p\eta\tilde\mu\tau+\tau\\
%		%0 & 1 & 1 & -Q_{4,2}-Q_{4,3}\end{pmatrix}. \\
%		Q = \begin{pmatrix}-\sum_{\ell\neq 1} Q_{1,\ell}&(d_j-1)\eta\tilde\mu\tau/n&(d_i-1)\eta\tilde\mu\tau/n&0\\
%			1 & -\sum_{\ell\neq 2} Q_{2,\ell} & 0 & ((d_i-1)\eta\tilde\mu\tau/n)+\tau\\
%			1 & 0 & -\sum_{\ell\neq 3} Q_{3,\ell} & ((d_j-1)\eta\tilde\mu\tau/n)+\tau\\
%			0 & 1 & 1 & -\sum_{\ell\neq 4} Q_{4,\ell}\end{pmatrix}. 
%	\end{align}
\begin{figure*}
	\begin{tcolorbox}[colback=lightgray]
		\normalsize
		Given: a connected graph $G=(V,E)$ with $V = \{1,\ldots,n\}$. Infection rate: $\tau$. Healing rate: 1. 
		\begin{enumerate}
			\item Compute 
			\begin{align}
			P(d) := \frac{\#\{i\in V:\text{deg}(i) = d\}}{n},\qquad \tilde P(d) = \frac{d\cdot\#\{i\in V:\text{deg}(i) = d\}}{\sum_i d_i}.
			\end{align}
			%			\item Let $D$ be the degree distribution. Compute 
			%			\begin{align}
			%				P(d) := \mathbb{P}(D=d),\qquad
			%				\tilde P(d) := d\cdot\mathbb{P}(D=d)/\mathbb{E}[D].
			%			\end{align}	
			\item For each pair of degrees $(d_i,d_j)$, define a matrix $Q := Q(\eta,\tilde\mu)$ by
			\begin{align}\label{eq:Qmatrix}
			Q = \begin{pmatrix}-\sum_{\ell\neq 1} Q_{1,\ell}&(d_j-1)\eta\tilde\mu\tau/n&(d_i-1)\eta\tilde\mu\tau/n&0\\
			1 & -\sum_{\ell\neq 2} Q_{2,\ell} & 0 & ((d_i-1)\eta\tilde\mu\tau/n)+\tau\\
			1 & 0 & -\sum_{\ell\neq 3} Q_{3,\ell} & ((d_j-1)\eta\tilde\mu\tau/n)+\tau\\
			0 & 1 & 1 & -\sum_{\ell\neq 4} Q_{4,\ell}\end{pmatrix}, 
			\end{align}
			and determine the solution $\mathbf{x} = \mathbf{x}(d_i,d_j)$ (as function of $\eta$ and $\tilde\mu$) of 
			\begin{align}\label{eq:corsystem}
			\begin{cases}\mathbf{x}^TQ =\mathbf{0}^T,\\ \mathbf{x}^T\mathbf{1} = 1,\end{cases} \qquad
			\mathbf{x}^T = \begin{pmatrix}x_1 & x_2& x_3& x_4\end{pmatrix},
			\end{align}
			where $\mathbf{0}$ and $\mathbf{1}$ are the all zero and all one vector in $\mathbb{R}^4$. 
			\item Determine $\eta$ and $\tilde \mu$ by solving the equations
			\begin{align}
			&\frac{\tilde\mu}{n} = \sum_{d_i}\sum_{d_j}P(d_i)\tilde P(d_j)\left(x_2(d_i,d_j)+x_4(d_i,d_j)\right),
			&\frac{\eta\tilde\mu}{n} = \sum_{d_i}\sum_{d_j}P(d_i)\tilde P(d_j)\frac{x_2(d_i,d_j)}{x_1(d_i,d_j)+x_2(d_i,d_j)}.
			\end{align}
			\item Estimate the infected fraction by 
			\begin{align}
			\frac{\mu}{n} = \sum_{d_i}\sum_{d_j}P(d_i)\tilde P(d_j)\left(x_3(d_i,d_j)+x_4(d_i,d_j)\right).
			\end{align}
		\end{enumerate} 
	\end{tcolorbox}\caption{Main result: Algorithm (quenched) to estimate the metastable infected fraction in a connected graph $G$ with $n$ nodes.}\label{fig:mainresult}
\end{figure*}

The matrix $Q$ has rank 3 and \eqref{eq:corsystem} has a unique solution, giving $x_1,\ldots,x_4$ as functions of $n$, $\tau$, $d_i$, $d_j$, $\eta$ and $\tilde\mu$.  In principle these functions can be determined exactly. We will not do this, but we will show that in this system $X_i$ and $X_j$ have positive correlation, which vanishes if the degrees go to infinity. This agrees with conjectured and simulated behaviour of the SIS process. 

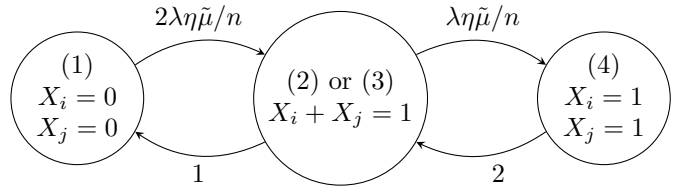
\begin{figure}
	\begin{center}
		\begin{tikzpicture}[->,>=stealth,scale=.7,auto,bend left]
		%\useasboundingbox (-1,-0.5) rectangle (3,1.8);
		\node[circle,draw,inner sep=-2pt,minimum width=5mm] (1) {\begin{tabular}{c}(1)\\$X_i=0$\\$X_j=0$\end{tabular}};
		\node[circle,draw,inner sep=-2pt,minimum width=5mm] (2) at ($ (1) +  (0:5) $) {\begin{tabular}{c}(2) or (3)\\$X_i+X_j=1$\end{tabular}};
		\node[circle,draw,inner sep=-2pt,minimum width=5mm] (3) at ($ (2) +  (0:5)$) {\begin{tabular}{c}(4)\\$X_i=1$\\$X_j=1$\end{tabular}};
		%\node[circle,draw,inner sep=-2pt,minimum width=5mm] (4) at ($ (3) + (150:5)$) {\begin{tabular}{c}(2)\\$X_i=0$\\$X_j=1$\end{tabular}};
		
		\path (1) edge[bend left] node[above]{$\displaystyle 2\lambda\eta\tilde\mu/n$} (2)
		(2) edge[bend left] node[above]{$\displaystyle\lambda\eta\tilde\mu/n$} (3)
		(2) edge[bend left] node[below]{1} (1)
		(3) edge[bend left] node{2} (2);
		\end{tikzpicture}
		\vspace{-.5cm}
	\end{center}
	\caption{Transition diagram for infectious status of $i$ and $j$ as $np\to\infty$}
	\label{fig:transitiondiagramasymptotics}		
\end{figure}

First we take a look at the Markov chain for $np\to\infty$. In this case, by the law of large numbers, $(d_i-1)/(np)\to 1$ and $(d_j-1)/(np)\to 1$ as $n\to\infty$. Since $\tau = \lambda/(np)$ is of order $(np)^{-1}$, all rates to the right are of constant order and asymptotically equal to $\lambda\eta\tilde\mu/n$. By symmetry, states (2) and (3) will then have the same stationary probability. The Markov chain therefore asymptotically simplifies to the one in Figure \ref{fig:transitiondiagramasymptotics}. Solving for the stationary distribution of this system, we can easily find $\mathbf{x}$ as well:
\begin{align}\label{eq:xasymp}
\mathbf{x}^T = \frac{1}{(1+c)^2}\cdot \left(1,c,c,c^2\right),\qquad\text{where\ }c = \frac{\lambda\eta\tilde\mu}{n}.
\end{align}
We note that from this solution it follows that nodes $i$ and $j$ become independent as the degrees go to $\infty$:
\begin{align}
\mathbb{P}(X_i=0)\cdot&\mathbb{P}(X_j=0) = (x_1+x_2)(x_1+x_3)\nonumber\\ &= \left(\frac{1+c}{(1+c)^2}\right)^2 = \mathbb{P}(X_i=X_j=0).
\end{align}
Now consider the more complicated Markov chain of Figure \ref{fig:transitiondiagram}. To avoid very awkward calculations, define
\begin{align}
a = \frac{(d_i-1)\eta\tilde\mu\tau}{n},\qquad b = \frac{(d_j-1)\eta\tilde\mu\tau}{n}. 
\end{align}
Since $\tau>0$, the rate from $X_i=0$ to $X_i=1$ is at least $a$. Therefore,
\[
\mathbb{P}(X_i=0) < \frac{1}{1+a},\qquad \mathbb{P}(X_j=0) < \frac{1}{1+b},
\]
so that in particular there exist constants $r_1,r_2<1$ for which $x_1+x_2= r_1/(1+a)$ and $x_1+x_3= r_2/(1+b)$.
Further, the detailed balance equation for state (1) yields $(a+b)x_1 = x_2+x_3$, and hence
\begin{align}
(2+a+b)x_1 = 2x_1+x_2+x_3 = \frac{r_1}{1+a}+\frac{r_2}{1+b}.
\end{align}
Solving this for $x_1=\mathbb{P}(X_i=X_j=0)$, we obtain
\begin{align}
\mathbb{P}(&X_i=X_j=0) = \frac{r_1(1+b)+r_2(1+a)}{(1+a)(1+b)(2+a+b)}\nonumber\\
&>\frac{r_1r_2}{(1+a)(1+b)} = \mathbb{P}(X_i=0)\cdot\mathbb{P}(X_j=0),
\end{align}
proving positive correlation for all choices of the parameters.

As noted, given the parameters $n, \tau, d_i, d_j$, we can determine the solution $\mathbf{x}$. However, this solution will still depend on the unknown $\eta$ and $\tilde\mu$. To solve for these variables, we need additional equations. 

So far, $i$ and $j$ have played the same role. But now we consider $j$ to be a random neighbor of a randomly chosen node $i$. This is only possible if $i$ has degree at least 1. In a connected graph, this is automatic. In an Erd\H{o}s-R\'enyi graph with small $p$, it means that $d_i$ has distribution $\text{Bin}(n-1,p)$, conditioned on being non-zero. Node $j$ has the size biased degree distribution $d_j-1\sim \text{Bin}(n-2,p)$, which is non-zero by definition. Consequently, $j$ has the same degree distribution as all neighbors of $i$ and $j$ (except $i$ itself). It therefore makes sense to approximate the expected fraction of time $j$ is infected by  the same $\tilde\mu/n$. Concretely, for each combination of $d_i$ and $d_j$, we solve the system \eqref{eq:corsystem} for $\mathbf{x} = \mathbf{x}(d_i,d_j)$. In particular we get a different solution 
\[\mathbb{P}(X_j=1)=x_2(d_i,d_j)+x_4(d_i,d_j)\]
for each combination $(d_i,d_j)$. Then we take a weighted average according to the degree distributions of $i$ and $j$. The resulting equation is
%	\begin{align}\label{eq:mutilde1}
%		\frac{\tilde\mu}{n} = \sum_{d_i=1}^{n-1}\sum_{d_j=1}^{n-1}P^{(\ge 1)}(n,p,d_i)\tilde P(n,p,d_j)\left(x_2(d_i,d_j)+x_4(d_i,d_j)\right),
%	\end{align}
\begin{align}\label{eq:mutilde1}
\frac{\tilde\mu}{n} = \sum_{d_i=1}^{n-1}\sum_{d_j=1}^{n-1}P^{(\ge 1)}(d_i)\tilde P(d_j)\left(x_2+x_4\right),
\end{align}
where $P^{(\ge 1)}(d_i)$ and $\tilde P(d_j)$ are the probabilities that $i$ and $j$ have degrees $d_i$ and $d_j$ respectively. That is, 
\begin{align*}
&P^{(\ge 1)}(d_i) = \mathbb{P}(D=d_i\mid D\geq 1),&&D\sim \text{Bin}(n-1,p),
\\
&\tilde P(d_j) = \mathbb{P}(\tilde D=d_j),&&\tilde D\sim 1+\text{Bin}(n-2,p). 
\end{align*}
Of course these probabilities can be replaced by suitable approximations for large $n$ or by graph-based estimates (see Section \ref{sec:numerics5}).

Another equation is obtained by considering conditional probabilities. Since $\mathbb{P}(X_j=1\mid X_i=0)$ hardly depends on the degree of $i$ (supported by Figure \ref{fig:condprob}), we can assume this conditional probability to be equal to $\eta\tilde\mu$. In terms of $\mathbf{x}$, we have 
\[
\mathbb{P}(X_j=1\mid X_i=0)=\frac{x_2}{x_1+x_2}.\] 
Proceeding along the same lines as above, we obtain 
\begin{align}\label{eq:eta}
\frac{\eta\tilde\mu}{n} &= \sum_{d_i=1}^{n-1}\sum_{d_j=1}^{n-1}P^{(\ge 1)}(d_i)\tilde P(d_j)\frac{x_2}{x_1+x_2}.
\end{align}
If all parameters of the process are given, the equations \eqref{eq:corsystem}, \eqref{eq:mutilde1} and \eqref{eq:eta} can be solved to find $\eta$ and $\tilde\mu$. Finding exact expressions is a tall order, but an iterative numerical procedure gives satisfactory results.

The solutions for $\eta$ and $\tilde\mu$ only depend (in a complicated way) on the process parameters $n$, $p$ and $\tau$. Once they have been determined, we can plug them into $\mathbf{x}$, which will then be a function of the same parameters and the degrees $d_i$ and $d_j$. This allows to compute an estimate for $\mu$ in the same fashion, by using that $\mathbb{P}(X_i=1) = x_3+x_4$ and taking a weighted average similar to \eqref{eq:mutilde1}. This time, we first pick a random node $i$. If it has degree 0, it will not contribute to $\mu$ (it quickly heals and never gets infected again). If it has degree $\geq 1$, we let $j$ be a random neighbor of $i$. The estimate for $\mu$ then is
\begin{align}\label{eq:mu1}
\frac{\mu}{n} = \sum_{d_i=1}^{n-1}\sum_{d_j=1}^{n-1}P(d_i)\tilde P(d_j)\left(x_3+x_4\right),
\end{align}
with 
\begin{align}
&P(d_i) = \mathbb{P}(D=d_i),&&D\sim \text{Bin}(n-1,p).
\end{align}
The subtle difference with \eqref{eq:mutilde1} and \eqref{eq:eta} is that now $i$ is not conditioned to have degree $\geq 1$, though still degree 0 does not contribute to the sum. 

The resulting equations for $\tilde\mu$ and $\mu$ can be seen as a more sophisticated version of the Heterogeneous Mean Field method (\eqref{eq:mutilde0} and \eqref{eq:mu0}). It takes size bias effects and correlations simultaneously into account. We now obtain a new annealed heuristic for estimating $\mu$ as a function of $n$, $p$ and $\tau$:

\begin{heuristic}[Annealed]\label{heuristic5}
	Let $G=(V,E)$ be an Erd\H{o}s-R\'enyi graph with parameters $n$ and $p$. Consider the SIS process on $G$ with infection rate $\tau$. For each $1\leq d_i,d_j\leq n-1$, let $\mathbf{x}(d_i,d_j)$ be the solution of \eqref{eq:corsystem}, which depends on the unknowns $\tilde\mu$ and $\eta$. Solve \eqref{eq:mutilde1} and \eqref{eq:eta} for $\tilde\mu$ and $\eta$ and substitute the solutions in $\mathbf{x}(d_i,d_j)$. Finally, take the weighted average \eqref{eq:mu1} to obtain an estimate for $\mu$.  
\end{heuristic}

This heuristic can easily be generalized to any connected graph, given its degree sequence (Figure \ref{fig:mainresult}). 

Before turning to the numerical results, we wrap up our discussion on asymptotics for $np\to\infty$. The degree distributions of $d_i$ and $d_j$ concentrate around their expectations, and $\mathbf{x}$ becomes independent of the degrees as in \eqref{eq:xasymp}. The equations \eqref{eq:mutilde1} and \eqref{eq:eta} become
\begin{align}
&\frac{\tilde\mu}{n} = x_2+x_4 = \frac{c+c^2}{(1+c)^2} = \frac{c}{1+c}\\
&\frac{\eta\tilde\mu}{n} = \frac{x_2}{x_1+x_2} = \frac{c/(1+c)^2}{(1+c)/(1+c)^2} = \frac{c}{1+c},
\end{align}
and solving them gives 
\begin{align}
%\tilde\mu = \frac{\lambda-1}{\lambda}n,\qquad \eta=1.
\tilde\mu = (1-\lambda^{-1})n,\qquad \eta=1.
\end{align}
Since $x_2=x_3$ in this case, we also get $\mu = \tilde\mu$. The conclusion is that according to Heuristic \ref{heuristic5}, both the size bias effect and the correlations vanish if the degrees go to infinity. This conclusion is consistent with simulations and intuition. Also note that the asymptotic estimate for $\mu$ agrees with Heuristic \ref{heuristic1}, which can be seen as a small sanity check. Finally, no solution exists for $\lambda<1$, reflecting the fact that the process has no metastable behaviour in this case.

\subsection{Quenched variants}

Heuristic \ref{heuristic5} can be reinforced if information about the graph is available. In particular, we can replace $P(d)$, $P^{(\geq 1)}(d)$ and $\tilde P(d)$ with graph-based estimates. We only have to do this for $d\geq 1$. We have the following quenched variants of Heuristic \ref{heuristic5}:

\addtocounter{heuristic}{-1}
\begin{heuristic}[Quenched variants]
	\begin{enumerate}[(a)]
		\item\label{heuristic5a} Estimate $p$ by the number of edges $m$ in the graph: $\hat p = \frac{2m}{n(n-1)}$.% and use $P(n,\hat p,d)$, $P^{(\ge 1)}(n,\hat p,d)$ and $\tilde P(n,\hat p,d)$.
		\item\label{heuristic5b} Estimate probabilities by true degree frequencies in the graph:
		\begin{align*}
		&P(d) = \frac{\#\{i:d_i=d\}}{n} = \frac1n\sum_{i=1}^n \mathbf{1}\{d_i=d\},\\
		&P^{(\ge 1)}(d) = \frac{\#\{i:d_i=d\}}{\#\{i:d_i\ge 1\}} = \frac{\sum_{i=1}^n \mathbf{1}\{d_i=d\}}{\sum_{i=1}^n \mathbf{1}\{d_i\ge 1\}},\\
		&\tilde P(d) = \frac1{\#\{i:d_i\ge 1\}}\sum_{i=1, d_i\ge 1}^n\frac1{d_i}\sum_{j\sim i}\mathbf{1}\{d_j=d\}.
		\end{align*}
		\item\label{heuristic5c} Estimate products of probabilities by true frequencies in the graph:
		\begin{align*}
		&P(d)\cdot\tilde P(e) %= \frac1n\sum_{i=1}^n\frac{\mathbf{1}\{d_i=d\}}{d_i}\sum_{j\sim i}\mathbf{1}\{d_j=e\}\\
		=\frac1n\sum_{i=1}^n\sum_{j\sim i}\frac{\mathbf{1}\{d_i=d,d_j=e\}}{d},\\
		&P^{(\geq 1)}(d)\cdot\tilde P(e) 
		=\frac{1}{\#\{i:d_i\geq 1\}}\sum_{i=1}^n\sum_{j\sim i}\frac{\mathbf{1}\{d_i=d,d_j=e\}}{d}.
		\end{align*}
	\end{enumerate}
\end{heuristic}  
As in NIMFA, one could even use the full adjacency matrix of the graph, but computations would be more complicated.

\section{Heuristic \ref{heuristic5}, numerical results}\label{sec:numerics5}

In this section we review the performance of annealed and quenched variants of Heuristic \ref{heuristic5}. For $p$ equal to the connectivity threshold $\log(n)/n$, Heuristics \ref{heuristic2} and \ref{heuristic3} are already very good (Figures \ref{fig:heur2sim} and \ref{fig:heuristics_boxplots}). Nevertheless, Heuristic \ref{heuristic5} seems to give a small improvement here. We will not discuss this in detail.

The real test case is the very sparse Erd\H{o}s-R\'enyi graph with $p$ significantly below the connectivity threshold. In this case all heuristics which do not take into account correlations fail to accurately estimate. This is the case for NIMFA and HMF, but in this sparse regime it also applies to Heuristics \ref{heuristic2} and \ref{heuristic3}. We will demonstrate how Heuristic \ref{heuristic5} can be used to get accurate estimations in this regime (Section \ref{sec:sparse}). 

Heuristic \ref{heuristic5} does not only give accurate estimates for the infected fraction of the population. As a bonus, it can also be exploited to obtain estimates for other quantities like correlations between nodes (Section \ref{sec:5other}).   

\subsection{Estimation in sparse graphs, the regime $p = c/n$}\label{sec:sparse}

The effect of correlations and size bias becomes more important when degrees in the graph are smaller. So far we mostly looked at Erd\H{o}s-R\'enyi graphs with $p$ close to or above the connectivity threshold. In these cases, variants of Heuristic \ref{heuristic2} still do reasonably well. Below the connectivity threshold, the picture changes. 

\begin{figure*}
	\includegraphics[width=0.5\columnwidth]{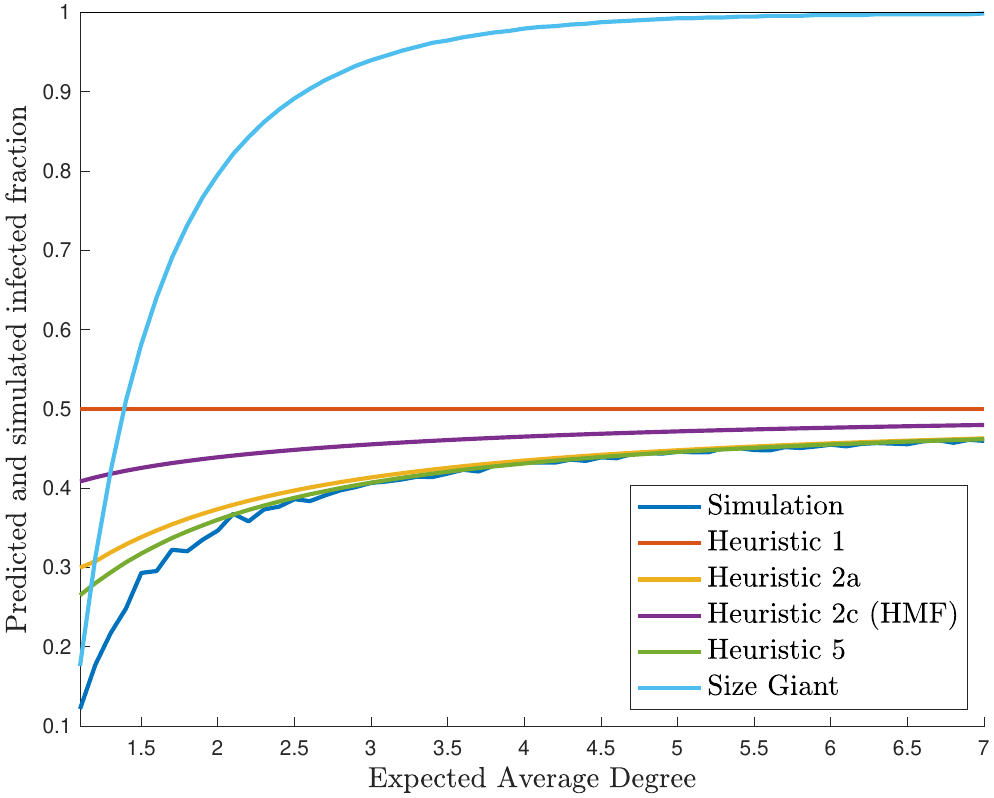}\includegraphics[width=.5\textwidth]{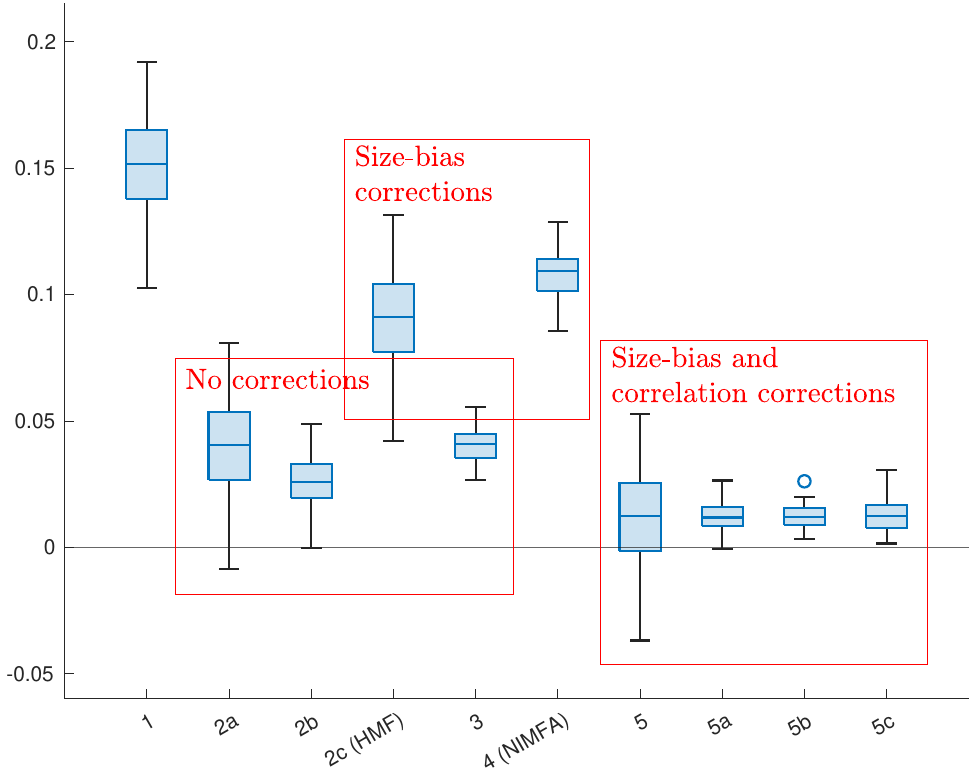}\caption{Sparse Erd\H{o}s-R\'enyi graphs ($n=1000$, $\lambda = 2$). Left: simulation compared to annealed heuristics. Right: estimation errors for all heuristics in Erd\H{o}s-R\'enyi graphs with expected average degree $np=2$.}\label{fig:simulationsparse2}
\end{figure*}

In the left plot of Figure \ref{fig:simulationsparse2}, we compare all annealed heuristics for Erd\H{o}s-R\'enyi graphs with $p=c/n$ and the average degree $c$ ranging from $1.1$ to $7$ $(\approx\log(n))$. For each parameter combination and heuristic, an average of simulations is compared with the corresponding estimates. In the plot on the right, we zoom in on the errors for 100 replications of the case $c=2$.

%\begin{figure}
%	\begin{center}
%\includegraphics[width=.8\textwidth]{Heuristics_Simulation_Averages.pdf}\caption{Sparse Erd\H{o}s-R\'enyi graphs: average of simulations compared to predictions by all different heuristics.}\label{fig:simulationsparse}
%\end{center}
%\end{figure}

%The annealed and quenched versions of Heuristic \ref{heuristic5} give good results for (almost) connected Erd\H{o}s-R\'enyi graphs, i.e. for edge probability $p>\log(n)/n$. 

For $p = c/n$, $c>1$, the graph disconnects and there will be a unique giant component. This giant component has size about $y\cdot n$, with $y$ the largest solution of $1-y=e^{-cy}$, see \cite{Hofstad17}. All other components will be much smaller, of size at most logarithmic in $n$. In these small components the infection quickly disappears. The metastable infected fraction of the population therefore will be less than the size of the giant component. The simulation in Figure \ref{fig:simulationsparse2} confirms this, but none of the heuristics captures this effect. 

\begin{figure}
	\includegraphics[width=\columnwidth]{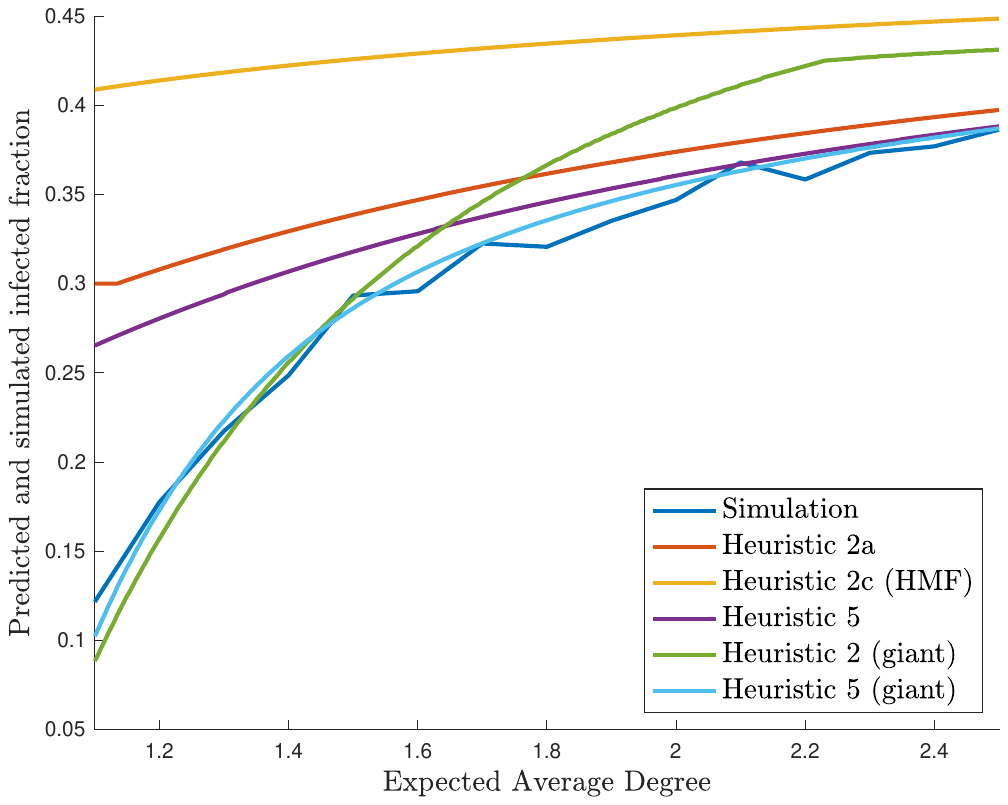}\caption{Most heuristics are wrong in very sparse Erd\H{o}s-R\'enyi graphs ($n=1000$, $\lambda = 2$). One should take the degree distribution of the giant component and apply size bias and correlation corrections (light blue).}\label{fig:sparse}
\end{figure}

Our implementations of Heuristic \ref{heuristic5} so far do not account for these properties of subcritical Erd\H{o}s-R\'enyi graphs. For the degree distribution they just take the binomials which apply to the whole graph. However, for small $c$, one should restrict the analysis to the giant and use the corresponding degree distributions. The degree distributions in the entire graph, in the giant component and their size biased versions are all different in the sparse regime. Only when the degrees go to infinity, these differences will be negligible. The calculation of these distributions is elementary. Heuristic \ref{heuristic5} can then be used to estimate the infected fraction within the giant component. After rescaling by the expected size of the giant component, we obtain annealed estimates for $\mu/n$ in the Erd\H{o}s-Renyi graph. Note that this procedure still only uses the parameters, no graph information is needed.

Figure \ref{fig:sparse} shows that this gives pretty accurate results (see also the red curves in Figure \ref{fig:Heuristics_Simulation}). It also shows that it is insufficient to merely use the correct degree distributions for the giant, apply Heuristic \ref{heuristic2} and then rescale by the size of the giant component. Correlations and size bias have to be taken into account as well. 

In a similar fashion quenched heuristics can be applied to the giant component. An illustration is given in Figure \ref{fig:Heuristics_Simulation}, where the quenched method only uses the sum of the degrees as in Heuristic \ref{heuristic5}\ref{heuristic5a} (so not the actual degrees, not the actual size of the giant component). These quenched versions also turn out to give very accurate estimates.

The analysis of the giant component serves as a test case for graphs with other degree distributions. Since our methods work quite well in the giant component, we expect them to apply to graphs with other degree distributions as well. Which features of the graph are essential to make the methods work requires further research. We expect that the methods are less suitable for graphs with a high variation in degrees (like power-law degree distributions) and for graphs which locally do not look like a tree (like grids). Graphs for which we do expect good results include the random regular graph.

\subsection{Estimating per-node infected fractions, correlations and the total variance}\label{sec:5other}

The ideas behind Heuristic \ref{heuristic5} give more information than just an estimate for $\mu$.  After solving all equations, we know $\mathbf{x}(d_i,d_j)$ for all combinations of $d_i$ and $d_j$. This means that we can make degree-dependent estimates. For instance, to estimate the fraction of time a random node of degree $d$ is infected, take $d_i = d$ and average over $d_j$ (cf. \eqref{eq:mu1}). Similarly, we can estimate conditional probabilities. 

\begin{figure}
	\begin{center}
		\includegraphics[width=\columnwidth]{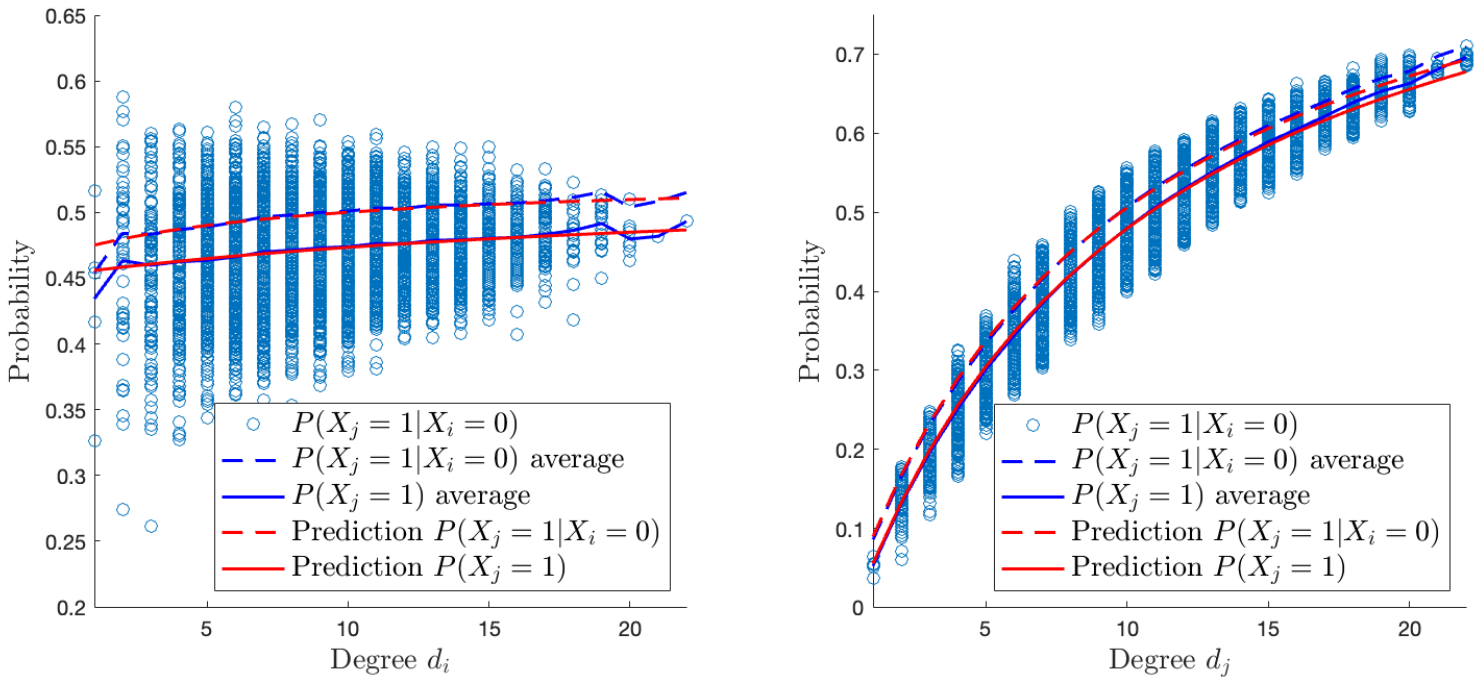}\caption{Infection probabilities and conditional infection probabilities for pairs $(i,j)$ with estimates (ER graph $n=10^4$, $p = \log(n)/n$, $\lambda = 2$). Left: averaged over neighbors of $i$ and as function of $d_i$. Right: as function of $d_j$.}\label{fig:predictcondprob}
	\end{center}
\end{figure}

For illustration, see Figure \ref{fig:predictcondprob}. The simulation is essentially the same as Figure \ref{fig:condprob}, but now we can estimate these quantities as follows. Let $i$ and $j$ be neighboring nodes. First we solve for $\mathbf{x}(d_i,d_j)$ as in Heuristic \ref{heuristic5}. An approximation for $\mathbb{P}(X_j=1\mid X_i=0)$ is $x_2/(x_1+x_2)$. Fixing $d_i$ and taking a weighted average over all values of $d_j$ gives an estimate for $\mathbb{P}(X_j=1\mid X_i=0)$ as function of $d_i$. Similarly, we get estimates as function of $d_j$. These estimates are as in \eqref{eq:eta}, but now only summing over $d_j$ or only over $d_i$ respectively. For the unconditional probabilities $\mathbb{P}(X_j=1)$, the procedure is analogous, but now using \eqref{eq:mutilde1}. In fact, by looking at pairs, we are estimating a version of a size biased distribution here. Similar methods can be used to estimate the infected fraction of time for a random node with degree $d_i$ using \eqref{eq:mu1} and only summing over $d_j$.

\begin{figure}
	\includegraphics[width=\columnwidth]{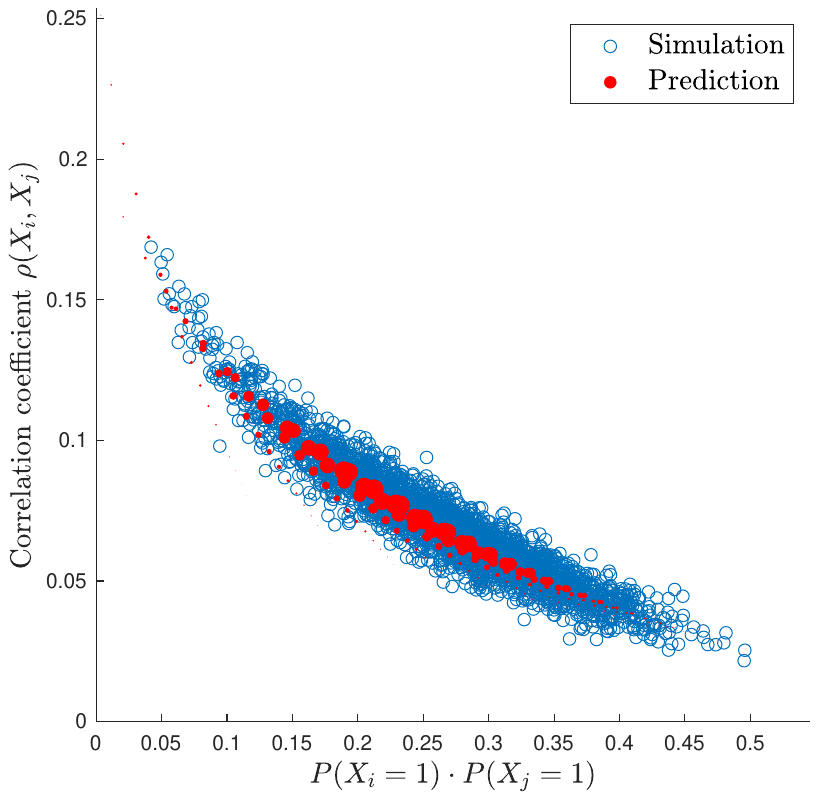}\caption{Simulated and estimated correlations. (ER graph $n=10^4$, $p = \log(n)/n$, $\lambda = 2$)}\label{fig:correlation_predict}
\end{figure}

Another feature of interest are correlations between neighbors. For an edge $(i,j)$ with degrees $d_i$ and $d_j$, we can solve for $\mathbf{x}(d_i,d_j)$, estimate the
(co)variances cf. \eqref{eq:covar} by
\begin{align}\label{eq:corpred}
\text{Cov}(X_i,X_j) &= x_4-(x_3+x_4)\cdot(x_2+x_4),\nonumber\\
\text{Var}(X_i) &= (x_3+x_4)(1-(x_3+x_4)),\nonumber\\
\text{Var}(X_j) &= (x_2+x_4)(1-(x_2+x_4)),
\end{align}
and then the correlation coefficient $\rho(X_i,X_j)$ of $i$ and $j$ by \eqref{eq:correlation}.
%		\begin{align}\label{eq:corpred}
%			\rho(X_i,X_j) \approx \frac{x_4-(x_3+x_4)\cdot(x_2+x_4)}{\sqrt{(x_3+x_4)(1-(x_3+x_4))\cdot (x_2+x_4)(1-(x_2+x_4))}},
%		\end{align} 
%		based on  
Plotting this against
\begin{align}\label{eq:bothinfpred}
\mathbb{P}(X_i=1)\cdot\mathbb{P}(X_j=1) \approx (x_3+x_4)\cdot(x_2+x_4)
\end{align}
gives Figure \ref{fig:correlation_predict}. The left hand sides of \eqref{eq:corpred} and \eqref{eq:bothinfpred} are simulated, the right hand sides are estimations. For each combination of degrees $d_i$ and $d_j$, there is one estimated point, plotted as a red marker. The size of the marker is proportional to the probability of the degree pair $(d_i,d_j)$. We see that correlations are maybe slightly underestimated for the less likely degree combinations. This could be explained by the fact that Heuristic \ref{heuristic5} ignores correlations at distance 2 and larger. 

One could wonder if we can use these correlation estimates to obtain estimates for the variance of the number of infected nodes. We have 
\begin{align}\label{eq:totalvariance}
\text{Var}\left(\sum_{i=1}^nX_i\right) = \sum_{i=1}^n\text{Var}(X_i) +2\sum_{i\neq j}\text{Cov}(X_i,X_j).   
\end{align}
For instance, in the complete graph expectation and variances are known: 
\begin{align*}
&\mathbb{E}[X_i] = 1-\frac1\lambda, \text{Var}(X_i)=\frac1\lambda(1-\frac1\lambda),\\ &\text{Var}\left(\sum X_i\right) =\frac n\lambda.
\end{align*}
Substituting in \eqref{eq:totalvariance} gives
\begin{align}
%\frac{n}{\lambda} = \frac{n}{\lambda}(1-\frac 1 \lambda)+n(n-1) \text{Cov}(X_1,X_2),
n\lambda^{-1} = n\lambda^{-1}(1-\lambda^{-1})+n(n-1) \text{Cov}(X_1,X_2),
\end{align}
So we can compute the covariances as well to find
$
\text{Cov}(X_1,X_2)\sim 1/(n\lambda^2).
$ This means that pair covariances are small compared to node variances, but the sum of covariances significantly contributes to the total variance.

In the Erd\H{o}s-R\'enyi graph, we can estimate $\mu = \sum \mathbb{E}[X_i]$ as in \eqref{eq:mu1}. Similarly, we estimate the sum of variances \[\sum_i \text{Var}(X_i) = \sum_i \mathbb{E}[X_i^2]-\mathbb{E}[X_i]^2\] by taking a weighted average of degree-dependent estimates. 

For covariances of neighbors, we have degree-dependent estimates $\text{Cov}(d_i,d_j)$ as well in \eqref{eq:corpred}. This can be used to estimate $2\sum_{i\sim j}\text{Cov}(X_i,X_j)$ by
\begin{align}
pn(n-1)\sum_{d_i=1}^{n-1}\sum_{d_j=1}^{n-1}\tilde P(d_i)\tilde P(d_j)\text{Cov}(d_i,d_j).
\end{align}
This is twice the expected number of edges, multiplied by the estimated covariance for a randomly chosen edge. A subtlety to be noted here is the weighting: if we pick a random edge, then both endpoints have the size biased degree distribution. 

\begin{figure}
	\includegraphics[width=\columnwidth]{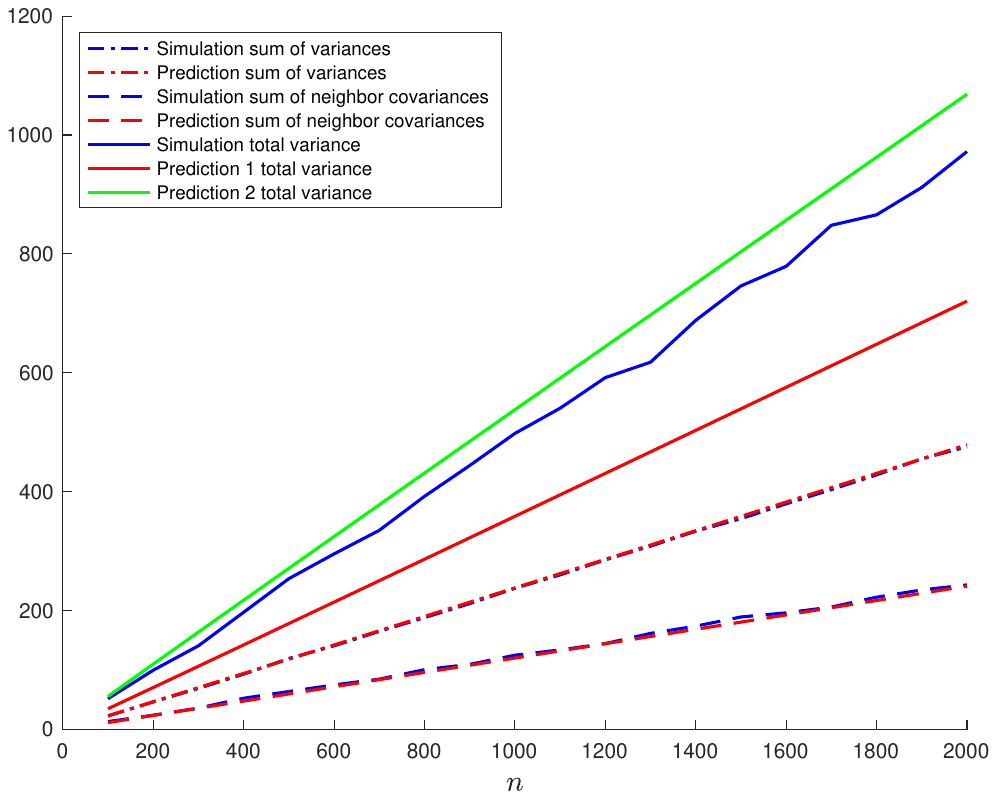}\caption{Simulations and estimates of variances and covariances. (ER graph $p = \log(n)/n$, $\lambda = 2$)}\label{fig:variance}
\end{figure}

What is still missing to estimate the total variance in \eqref{eq:totalvariance} are covariances for pairs of non-neighboring nodes. Figure \ref{fig:variance} shows simulations and estimates for the terms 
\begin{align}
\sum_i\text{Var}(X_i),\qquad 2\sum_{i\sim j}\text{Cov}(X_i,X_j). 
\end{align}
Both terms are estimated quite accurately. However, the sum of these two (solid red in the figure) turns out to be a bad estimate for the total variance, which was simulated as well. This means that we can not ignore correlations between nodes at mutual distance 2 or larger. The covariances $\text{Cov}(X_i,X_j)$ for $i\not\sim j$ are much smaller (cf. Figure \ref{distance_corr}), but the number of pairs is much larger, making the sum a significant contribution to the total variance. Since Heuristic \ref{heuristic5} is based on analysis of direct neighbors, it falls short to estimate the total variance.

There is however another way to estimate the total variance. In equilibrium, the total healing rate is equal to $\mu n$. By definition of equilibrium, this is equal to the total infection rate. For a uniformly at random selected set of $\mu n$ nodes, we expect $p\mu(1-\mu)n^2$ edges to its complement. The set of infected nodes is \emph{not} a uniform selection (e.g. high degrees are overrepresented). Still, if the number of infected $k$ is close to equilibrium, we can assume that the number of edges to the complement is proportional $k(n-k)$. In equilibrium ($k=\mu n$), the total infection rate is $\tilde p\tau\mu(1-\mu)n^2$, where $\tilde p$ is a conditional edge probability given that we consider a healthy and an infected node. This gives the equilibrium equation 
\begin{align}
\mu n = \tilde p\tau\mu(1-\mu)n^2,
\end{align}
so that $\tilde p\tau$ is equal to the reciprocal of the number $(1-\mu)n$ of healthy nodes in equilibrium. Note that this is consistent with the complete graph, when $p=\tilde p = 1$.

Now suppose the number of infected deviates a bit from equilibrium and is equal to $k = \mu n+d$. The ratio of total infection and total healing rate is then given by
\begin{align}
\frac{\tilde p\tau (\mu n+d)((1-\mu)n-d)}{\mu n+d} = 1-\frac{d}{(1-\mu)n}.
\end{align}
In this formula, we see a drift towards the equilibrium $\mu n$. This type of drift is known to lead to a Gaussian metastable distribution with variance $(1-\mu)n$, see \cite{ACD23}. Since we have an estimate for $\mu$, we have an estimate for the variance as well, which is added to the plot in Figure \ref{fig:variance} (solid green). This improves the estimate based on Heuristic \ref{heuristic5}, but is still biased. A better understanding of correlations of non-neighboring nodes will be needed to get more accurate variance estimates.

\bibliographystyle{plain} % We choose the "plain" reference style
%\bibliography{/Users/henkdon/Library/CloudStorage/Dropbox/Onderzoek/Susceptible_Infected_Susceptible/contactprocess}

\bibliography{contactprocessref}

\end{document}